\documentclass[prf, amsmath,amssymb,twocolumn,longbibliography]{revtex4-1}

\usepackage{times}
\usepackage{graphicx}
\usepackage{bm}
\usepackage{changes}

\usepackage{color}

\definecolor{pink}{rgb}{1,1,0} 
\definecolor{red}{rgb}{1,0,0}
\definecolor{yellow}{rgb}{1,1,0}
\definecolor{orange}{rgb}{1,0.5,0}
\definecolor{white}{rgb}{1,1,1}
\definecolor{blue}{rgb}{0,0,1}

\begin{document}
\title{Anisotropic finite-time singularity in the three-dimensional axisymmetric Euler equation with a swirl.}
\author{Sergio Rica}
\affiliation{
Instituto de F\'isica, Pontificia Universidad Cat\'olica de Chile, Avenida Vicu\~na Mackenna 4860, Santiago, Chile.
}

\begin{abstract}
The search of finite-time singularity solutions of Euler equations is considered for the case of an incompressible and inviscid fluid. Under the assumption that a finite-time blow-up solution may be spatially anisotropic as time goes by such that the flow contracts more rapidly into one direction than into the other, it can be shown that the dynamics of an axially symmetric flow with swirl may be approximated to a simpler hyperbolic system. By using the method of characteristics, it can be shown that generically the velocity flow exhibits multi-valued solutions appearing on a rim at a finite distance from the axis of rotation which displays a singular behavior in the radial derivatives of velocities. Moreover, the general solution shows a genuine blow-up which is also discussed. This singularity is generic for a vast number of smooth  finite-energy initial conditions and is characterized by a local singular behavior of velocity gradients and accelerations.
\end{abstract}

\maketitle

\section{Introduction}\label{Sec:Intro}

 Despite more than 250 years of history, a general understanding properties of solutions of Euler equations remains as an open problem. In particular the so-called regularity problem or the possible existence of finite-time singularity solutions~: Does a smooth initial condition for the velocity field remains regular for all times as the velocity flow evolves accordingly with Euler equations for an inviscid and incompressible fluid~?  By smooth we mean that the initial condition is differentiable everywhere and of finite energy (see the definition equation (\ref{eq:Energy}) below) which is one of the invariants of the dynamics. 
Although Euler's (and Navier-Stokes') equations look simple, these are hard to solve partial differential equations because they are of nonlinear and non-local character, therefore the raised question remains elusive, as well as, for instance the turbulent motion in Navier-Stokes does.

The search of singularities of solutions of  Euler equation for fluids in three space dimensions is not new, early attempts go back to the early 20th century \cite{Lichtenstein,Gunther}. In the '30s, Leray \cite{leray1934} suggested the possibility of a self-similar solution of Navier-Stokes equations in which the velocity field scales as a power law in time. Following Leray's endeavor, Pomeau  \cite{yves1995,yves2018,yves2019,martine19}, Chae \cite{Chae2007a,Chae2007b,Chae2010}, and others have revisited a possible existence of point-like spatio-temporal singularities. Nevertheless, Leray's explicit self-similar equations for singularities lead to a very challenging problem that is far from being fully understood.  Unfortunately, this approach  has not been successful in providing an explicit example of such type of singularity.

Theoretical physics offers normally the possibility of existence of singular behaviors which may be regarded as internal paradoxes. Some examples of singular solutions in partial differential equations are: the electric field created by a point-like charge, the magnetic field induced by a current in a wire, a vortex ring in an incompressible fluid \cite{lamb1895hydrodynamics,Saffman1995vortex}, 
the space-time singularities in general relativity \cite{Penrose1965}, to name a few examples.  

In the context of fluid motion, a geometric approach based on singular structures as vortex sheets successfully manifest a singular behavior \cite{Moore1979}. Similarly, Siggia suggested that interactions of vortex filaments may result in this kind of self-similar reconnection \cite{Siggia85,PumirSiggia87}. However, an asymptotic limit for a solution of Euler equation into singular manifolds is not entirely satisfactorily because of the lack of any intrinsic length in Euler equations.

In the last forty years, as a consequence of improvement, in computer technology, the question of regularity regained interest from a numerical and theoretical approach. Both physicists and mathematicians made an enormous effort in the search of possible singular solutions of Euler equations. To mention a few, the Taylor-Green initial condition was considered in Refs. \cite{Orszag1980,Brachet1983,Brachet92}, antiparallel vortex tubes \cite{Kerr93,Kerr2005}, and, high-symmetric Kida flow \cite{BoratavPelz94}.
The author refers to the review article by Gibbon \cite{Gibbon08} who carefully reviews the most notable progress including a summary table with expectations of existence, or not, of finite-time blow-up solutions of Euler, as well as Navier-Stokes, equations. More recently, Luo and Hou \cite{Luo2014} provided numerical evidence of the existence of a finite-time blow-up for the axially symmetric Euler equations at the outer boundary of the domain and Barkley \cite{Barkley2020} analyses this singularity from a hydrodynamical aspect showing that the formation of the singularity is driven by the wall. Lastly, Elgindi and Jeong prove the existence of a finite-time singularity of axially symmetric Euler equations in a ``hour-glass" like domain excluding the axis of rotation \cite{Elgindi2019}. The question of the existence of such singularities in the whole space remained as an open problem until 2021, when Elgindi \cite{Elgindi2021annals} showed how, under certain assumptions, the non-local contribution of vorticity can be simplified in the case of an axisymmetric flow without swirl exhibiting a self-similar blow-up in finite-time.


 In this article, we consider the possibility that due to fluid stretching the flow could destroy the spatial isotropy, modifying the temporal evolution of the flow, ruled by Euler equations, and, spontaneously, the dynamics generates a space anisotropy inducing a flow that may shrink more rapidly in one direction than into the other. 
 Indeed, this process is supported by the numerical work of Kerr (see Fig, 2 \& 4 of Ref. \cite{Kerr93}), as well as, by some recent work by Brenner, Hormoz, and Pumir \cite{PumirPRF2016}.  
Under this assumption, we show that the axi-symmetric flow may be approximated to a hyperbolic system which is solved using the method of characteristic. It is shown that, generically the radial and swirl velocities become multi-valued functions at a rim at a finite distance from the axis of rotation exhibiting a singularity of some velocity derivatives. 

 The current singularity appears to be different from the one found recently by Elgindi~\cite{Elgindi2021annals}. In the current paper, the singularity is a consequence of the advective structure of the flow dynamics together with the swirl and an adequate initial condition for the velocity flow. The advection effect makes the velocity flow to be a multivalued function at some time $t_c$, further $\frac{\partial v_r}{\partial r} \sim 1/(t_c-t)$. Contrarily, in Ref.~\cite{Elgindi2021annals}, as a consequence of a property exhibited by the axisymmetric flow without swirl, the advective term, ${\bm v}\cdot {\bm \nabla}$, can be discarded in the vorticity equation (eqn. (\ref{eq:Helmholtz}) below), while the non-locality (a simplified Biot-Savart integral) exhibits a genuine finite-time singularity, similar to the one studied by Constantin, Lax, and Majda \cite{ConstantinLaxMajda1985}, and De Gregorio \cite{DeGregorio1990,DeGregorio1996} in the 80s and 90s.

A better knowledge of the nature of solutions of Euler or Navier-Stokes equations may pave the path from the original Leray's idea  for approaching the problem of turbulence \cite{yves2018,yves2019}. Indeed, based on the self-focusing non-linear Schr\"odinger equation, in collaboration with C. Josserand and Y. Pomeau, we suggested a singularity-mediated turbulent scenario of nature for the observed intermittency in fully developed turbulence~\cite{PRFJoss2020}.

The paper is organized as follows, Section \ref{Sec:Euler}  introduces Euler equations and their basic properties~: symmetries and conserved laws. Further, the Leray finite-time singularity approach is briefly reviewed. In Section \ref{Sec:AxiSymmetric}, the equations for an axially symmetric flow with swirl are introduced. In this situation, a tridimensional velocity field plus pressure is reduced to a coupled system of two partial differential equations. Next, the main assumption of the paper allows to simplify the axi-symmetric flow making it possible an explicit solution.
Section \ref{Sec:Singular} presents the main result of the paper in which, via the method of characteristic, we show that the simplified anisotropic model  shows  two consecutive singularities in time. A primary singularity arises as the radial and swirl velocities become multivalued functions in the Eulerian description. This singularity, is formally followed by a later singularity inside the multi-valued domain. 
Section \ref{Sec:Discussion} concludes with a general discussion and future perspectives.  Appendix  \ref{Sec:Mapping} shows a qualitative point of view  by employing tools taken from the Hamiltonian dynamical system theory, and Appendix \ref{App:InfiniteEnergyExample} shows a particular infinite energy solution exhibiting a finite-time singular behavior in the axi-symmetric reduced model.

\section{Euler equations.} \label{Sec:Euler} 
Euler equations for an inviscid and incompressible fluid read:
\begin{eqnarray}
\frac{\partial }{\partial t}  {\bm v} ( {\bm x},t) + { \bm v} \cdot {\bm \nabla}  \, \, { \bm v}   &= &- {\bm \nabla} p , \label{eq:Euler} \\
{\bm \nabla} \cdot { \bm v}   &= &0. 
\label{eq:EulerContinuity}
\end{eqnarray}
These equations are complemented by  boundary conditions plus the initial flow velocity.
For the boundary conditions, as we precise later on, we assume that the velocity field decreases at infinity such that $\lim_{r\to \infty} |{\bm v}|^2 r^3 \to 0 $. 
Additionally to equations (\ref{eq:Euler}) and  (\ref{eq:EulerContinuity}), one needs a divergence free initial condition which reads:
\begin{eqnarray}
 {\bm v} ( {\bm x},t=0)   &= & {\bm v} _0( {\bm x})  , \nonumber\\ 
{\bm \nabla} \cdot { \bm v}_0   &= &0. 
\label{eq:EulerInit}
\end{eqnarray}

Equivalently, taking the curl of (\ref{eq:Euler}) eliminates the pressure term. This procedure gives an equation for the vorticity field:
\begin{eqnarray}
{\bm \omega} &=& {\bm \nabla} \times {\bm v} . \label{eq:vorticity}
\end{eqnarray}

The vorticity (or the Helmholtz) equation reads:
\begin{eqnarray}
\partial_t {\bm \omega} +{\bm v} \cdot  {\bm \nabla }  {\bm \omega}   &=&{\bm \omega} \cdot  {\bm \nabla }  {\bm v} ,
\label{eq:Helmholtz}
\end{eqnarray}
The right hand side in (\ref{eq:Helmholtz}) represents vorticity stretching: everywhere in the space, locally, the vorticity experiences a growth in one direction and a contraction into the other direction.

From a more taxonomic point of view, Euler equations are a set of nonlinear and nonlocal partial differential equations. The non-locality comes from the pressure term, $p( {\bm x},t)$, the which, after taken the divergence of (\ref{eq:Euler}) follows as a solution of a Poisson equation: 
\begin{eqnarray}
 {\bm \nabla}^2 p    &= &- 
  \partial_{ik} (v_i  v_k)\equiv - 
   \partial_i {  v_k}  \partial_k{  v_i}  . 
    \label{eq:Poisson} 
\end{eqnarray}
Here repeated indices stand for a sum as in Einstein's convention.
Thus, the pressure contribution becomes a non-local functional of the right-hand side of (\ref{eq:Poisson}).
Similarly, Helmholtz equation (\ref{eq:Helmholtz}) is also a nonlinear and nonlocal partial differential equation:  the velocity in (\ref{eq:Helmholtz}) is a nonlocal functional of vorticity (\ref{eq:vorticity}) as follows from the Biot-Savart law:

\begin{eqnarray}
 {\bm v}({\bm x})= \frac{1}{4\pi} \int {\bm \omega}({\bm x}') \times  \frac{ ({\bm x}-{\bm x}' )}{|{\bm x}-{\bm x}'|^3} \,  d{\bm x}' .
    \label{eq:BiotSavart} 
\end{eqnarray}

Euler equations, as well as Helmholtz equations, posses a  number of symmetries and conserved quantities. Among them in can be mentioned: {\it Space-time translational symmetry}, {\it  Time reversibility},  {\it  Rotational invariance}, {\it Galilean invariance}, and, {\it Scale invariance}. The last symmetry tells us: if $\forall \, {\ell} \in \mathbb{R}\, \& \, \tau \in \mathbb{R}$, and if   ${\bm v}({\bm x},t) $ and ${ p}({\bm x},t) $ are solutions of  (\ref{eq:Euler}) and  (\ref{eq:EulerContinuity}), then, $\frac{\ell}{\tau} {\bm v}( {\bm x}/\ell,t/\tau) $ and $ \frac{\ell^2}{\tau^2} { p}( {\bm x}/\ell,t/\tau) $, are also solutions of (\ref{eq:Euler}) and  (\ref{eq:EulerContinuity}). This scale invariance symmetry is at the basis of Leray's self-similar solutions that it is discussed later on.

Some conserved quantities in Euler equations are: the {\it Kinetic Energy}, {\it Linear momentum}, {\it total vorticity}, {\it Circulation conservation}, {\it Helicity conservation}, among others. We refer the reader to relevant textbooks for a deeper and additional review on conserved quantities \cite{lamb1895hydrodynamics,landau59,majda_bertozzi_2001}.

The kinetic energy reads
\begin{eqnarray}
{\mathcal E} = \frac{1}{2} \int_{\mathbb{R}^3}  | {\bm v}({\bm x},t) |^2 \, d {\bm x}, \label{eq:Energy}
\end{eqnarray}
 then, by (\ref{eq:Euler}) and  (\ref{eq:EulerContinuity}) it follows the conservation of the energy (\ref{eq:Energy}), {\it i.e.} $$\frac{d {\mathcal E}}{dt}=0.$$ In the following, excepting in the Appendix  \ref{App:InfiniteEnergyExample}, we restrict the discussion to finite energy flows,  therefore the initial condition must satisfy 
 $$ \int_{\mathbb{R}^3}   | {\bm v}_0({\bm x}) |^2 \, d {\bm x} < \infty.$$ It is worth mentioning, that infinite energy blow-up solutions could be found  in the existing literature \cite{Gibbon_2003}.

In 1934, Leray \cite{leray1934} suggested  a point-like singularity at the origin solution of equations of fluid dynamics (\ref{eq:Euler}) and  (\ref{eq:EulerContinuity}) of the form:
$$
{\bm  v} ({\bm x},t)  = \frac{1}{(t_c-t)^{1-\beta} }  {\bm V}\left(   \frac{\bm x}{ (t_c-t)^{\beta} } \right) . 
$$

Originally, Leray set $\beta=1/2$ as it was a parameter fixed by viscosity in Navier-Stokes equation.  In that situation, $\beta= 1/2$, Ne\v{c}as, R\r{u}\v{z}i\v{c}ka, and \v{S}ver\'ak have shown that the only solution for ${\bm V}(\cdot)$ satisfying the Navier-Stokes-Leray equation is ${\bm V}=0$  \cite{Necas1996}. However, the absence of viscosity together with scale invariance symmetry leaves free any possible relation between length and time, thus, {\it a priori}, $\beta$ is not fixed by dimensional analysis. We underline that a given initial value of energy or circulation may set a characteristic length,  {\it e.g.} $\beta = 2/5$ characterizes a flow in which the initial energy, $\mathcal E$, fixes the scales; $\beta=1/2$ corresponds to a scale fixed by circulation. Different exponents have been considered by Pomeau {\it et al.}~\cite{yves1995,yves2018,yves2019,martine19} and by Chae \cite{Chae2007a,Chae2007b,Chae2010} in a series of papers. Despite the efforts,  a satisfactory solution of the resulting self-similar Euler-Leray equation, satisfying the right boundary conditions, has became a very challenging problem, perhaps harder than the original time-dependent  Euler equations  (\ref{eq:Euler}) and  (\ref{eq:EulerContinuity}). 

Until now, the still unknown Leray's singularities have been supposed to be isotropic. That is, presumably the flow evolves independently of initial fluctuations toward a similar scaling of all coordinates in time, in other words, all principal axis scale in time with the same rate.
Nevertheless, there is no reason to exclude the possibility that the self-similar velocity field may scale in an anisotropic way in time. It is plausible that initial fluctuations may 
be amplified in one direction more than in another. 
This process is supported by numerical work of Refs. \cite{Kerr93} and \cite{PumirPRF2016}. Indeed, it has been suggested that vorticity stretching  (the right hand side of (\ref{eq:Helmholtz})) deforms vorticity to the extent that an initial elliptic vortex distribution is deformed in such a way that it becomes a vortex sheet as time evolves. Moreover, a simple argument based on a generic feature of the strain tensor, $ \partial_k v_i + \partial_i v_k$, indicates the existence of a dynamical ``shrink'' of, at least, in one coordinate: because of incompressibility, this symmetric tensor is locally traceless at all points in the domain, therefore, everywhere, at least one eigenvalue must be necessarily negative, and another must be positive.

By following an idea by Kasner \cite{Kasner1921c} for an anisotropic scaling of the space-time metric in General Relativity it is suggested  the following anisotropic self-similar velocity field: 
 \begin{widetext}
\begin{equation}
{ v}_i({\bm x},t)  = \frac{1}{(t_c-t)^{1-\beta_i} }  {V}_i\left(   \frac{x_1}{ (t_c-t)^{\beta_1} },  \frac{x_2}{ (t_c-t)^{\beta_2} },   \frac{x_3}{ (t_c-t)^{\beta_3} } \right) .
\label{eq:VelocityKasner}
\end{equation}
 \end{widetext}

The advantage of this dependence is that the divergence free condition (\ref{eq:EulerContinuity}) is preserved in the self-similar variables, and more important spatial gradients along with different  components are weaker than others, allowing a simpler systematic asymptotic analysis, as $t\to t_c^-$. Replacing (\ref{eq:VelocityKasner}) into Euler equations one gets that the pressure term $-{\bm \nabla } p$ becomes relevant just in the direction of the largest $\beta_i$. 
This observation opens the door to a new approach in the search of finite-time singularities of Euler equations, and is the general approach that will be pursued in a separate publication.

In the following section (Secc. \ref{Sec:AxiSymmetric}), it is applied the assumption of the anisotropic scaling to the simpler case of an axi-symmetric flow. In that situation, the assumption $\beta_1=\beta_2< \beta_3$ allows the approximation for the $i$-th component of the velocity~:
\begin{equation} \left| \frac{\partial   v_i}{\partial x} \right|\approx \left| \frac{\partial   v_i}{\partial y} \right|\ll \left| \frac{\partial   v_i}{\partial z} \right|,\label{eq:KasnerCriteria}
\end{equation}
simplifying the original 3D Euler equations. 

\section{The Axi-symmetric Flow with Swirl.}
\label{Sec:AxiSymmetric}

In the case of an axi-symmetric flow there is no dependence of any velocity on the angular variable, $\phi$, in cylindrical coordinates  \cite{Saffman1995vortex}. The velocity and vorticity fields read, respectively:
\begin{eqnarray}
 {\bm v} & = & \left( v_r(r,z,t), v_\phi(r,z,t), v_z(r,z,t)\right), \nonumber \label{eq:AxiSymVel}\\
 {\bm \omega} &= & \left( -\frac{\partial v_\phi}{\partial z}, \frac{\partial v_r}{\partial  z} - \frac{\partial  v_z}{\partial r}   ,\frac{1}{r} \frac{\partial (r v_\phi)}{\partial r} \right).\nonumber\label{eq:AxiSymOmega}
\end{eqnarray}
The incompressibility condition
$$ {\bm \nabla}\cdot {\bm v} =\frac{1}{r}  \frac{\partial (r v_r) }{\partial r} +  \frac{\partial v_z }{\partial z} =0,$$
introduces a stream function, $\psi(r,z)$, defined through the relationships:
\begin{eqnarray}
 v_r = -\frac{1}{r}  \frac{\partial \psi }{\partial z}, \quad & & \quad
  v_z = \frac{1}{r}  \frac{\partial \psi }{\partial r}. \label{eq:DefStream}
\end{eqnarray}
The axial vorticity component reads in terms of the stream function:
\begin{eqnarray}
 \omega_\phi &=& \frac{\partial v_r}{\partial  z} - \frac{\partial  v_z}{\partial r}=  -\frac{1}{r} \left(  \frac{\partial^2 \psi }{\partial r^2}  -\frac{1}{r}   \frac{\partial \psi }{\partial r}  +  \frac{\partial^2 \psi }{\partial z^2}  \right).
\label{eq:DefOmega}
\end{eqnarray}
Finally both, the vorticity, $\omega_\phi$, as well as the axial velocity, $v_\phi$, rule the self-contained system of partial differential equations  \cite{Saffman1995vortex}:

\begin{eqnarray}
\left(   \frac{\partial  }{\partial t}  + v_r  \frac{\partial  }{\partial r} + v_z  \frac{\partial  }{\partial z}  \right) \left(\frac{\omega_\phi}{r} \right) = \frac{1}{r^2}  \frac{\partial v_\phi^2 }{\partial z}, \label{eq:AxiSymmetricOmega} \\
\left(   \frac{\partial  }{\partial t}  + v_r  \frac{\partial  }{\partial r} + v_z  \frac{\partial  }{\partial z}  \right) \left({r v_\phi} \right) =0. \label{eq:AxiSymmetricGamma}
\end{eqnarray}
Equations  (\ref{eq:AxiSymmetricOmega}) and  (\ref{eq:AxiSymmetricGamma})  together with (\ref{eq:DefStream}) and (\ref{eq:DefOmega}) are formally a set of  two partial differential equations for the fields $\psi(r,z,t)$ and $v_\phi(r,z,t)$.
 The previous system was already numerically studied in the early 90s \cite{GrauerSideris91,PumirSiggia92} and, more recently, but in a finite domain in Refs. \cite{Luo2014,Barkley2020,Elgindi2019}.

The boundary conditions for  the  axi-symmetric flow with swirl at the axis of rotation, $r=0$, are such that 
\begin{eqnarray}
 v_r (r=0,z,t) &=&0,\nonumber\\ 
  v_\phi(r=0,z,t) &=&0, \label{eq:AxiSymmetricBCaxis}\\
   \left. \partial_rv_z (r,z,t) \right|_{r=0}&=& 0 . \nonumber
\end{eqnarray}
Moreover, $ v_r$ and $v_\phi$ are odd functions of the radial variable. The velocity boundary conditions (\ref{eq:AxiSymmetricBCaxis}) imply that the stream function $\psi$ is an even function in  the radial coordinate.  Thus \cite{Luo2014}: 
\begin{eqnarray}
\left. \frac{\partial^3 \psi} {\partial r^3} \right|_{r=0}= 0 , \label{eq:AxiSymmetricBCPsi}
\end{eqnarray}
and similarly in, all other derivatives of odd order.
Lastly, the velocity field decreases at infinite to ensure a finite energy flow.

\subsection{Anisotropic approximation.}

As previously mentioned, if the flow evolves in an anisotropic fashion, 
then the $z$-derivatives are usually larger than the radial ones. Therefore, the relevant approximation for equation (\ref{eq:DefOmega}) becomes:
$$  \omega_\phi \approx  -\frac{1}{r} \frac{\partial^2 \psi }{\partial z^2}, $$
or, in other terms: $ \left| \frac{\partial v_r }{\partial z}\right| \gg  \left| \frac{\partial v_z }{\partial r}\right| $.
This approximation allows us to simplify equations  (\ref{eq:AxiSymmetricOmega}) and  (\ref{eq:AxiSymmetricGamma}).

By replacing the above approximation and equations (\ref{eq:DefStream}) for  $v_r$, $v_z$ into equation (\ref{eq:AxiSymmetricOmega}) one obtains (after some simplification):
 \begin{widetext}
$$ \frac{\partial }{\partial z} \left( \frac{\partial^2\psi }{\partial t\partial z }   -     \frac{1}{  r}  \frac{\partial \psi }{\partial z }  \frac{\partial^2\psi }{\partial r\partial z }  +  \frac{1}{  r}  \frac{\partial \psi }{\partial r }   \frac{\partial^2\psi }{\partial z^2 }    +  \frac{1}{  r^2} \left(\frac{\partial \psi }{\partial z }  \right)^2  +  v_\phi^2  \right) =0.$$
 \end{widetext}
Whence, the expression inside-brackets must be independent of $z$:
$$\frac{\partial^2\psi }{\partial t\partial z }   -     \frac{1}{  r}  \frac{\partial \psi }{\partial z }  \frac{\partial^2\psi }{\partial r\partial z }  +  \frac{1}{  r}  \frac{\partial \psi }{\partial r }   \frac{\partial^2\psi }{\partial z^2 }    +  \frac{1}{  r^2} \left(\frac{\partial \psi }{\partial z }  \right)^2  +  v_\phi^2    =   I(r,t).$$
Here $I(r,t)$ is a function which may be solved by considering the asymptotic behavior of $ \frac{\partial \psi }{\partial z } $,  $\frac{\partial \psi }{\partial r}$  and $v_\phi$ as $z\to\pm\infty$. In the case of  finite energy solutions, we impose that both fields $v_\phi\to 0$ and $ \frac{\partial \psi }{\partial z } \to0$ as $z\to\pm\infty$, therefore the function $I(r,t)$ must be identically zero.
Then, by setting $I(r,t)=0$, and adding 
the swirl velocity equation (\ref{eq:AxiSymmetricGamma}), one gets the final coupled partial differential equations model:
\begin{eqnarray}
\frac{\partial^2\psi }{\partial t\partial z }   -     \frac{1}{  r}  \frac{\partial \psi }{\partial z }  \frac{\partial^2\psi }{\partial r\partial z }  +  \frac{1}{  r}  \frac{\partial \psi }{\partial r }   \frac{\partial^2\psi }{\partial z^2 }    +  \frac{1}{  r^2} \left(\frac{\partial \psi }{\partial z }  \right)^2  +  v_\phi^2    & = &   0,\nonumber\\ \label{eq:AxiSymmetricOmegaBis}\\ 
 \frac{\partial   }{\partial t} \left(r v_\phi\right) -\frac{1}{r}  \frac{\partial \psi }{\partial z}   \frac{\partial  }{\partial r} \left(r v_\phi\right)  +\frac{1}{r}  \frac{\partial \psi }{\partial r}  \frac{\partial  }{\partial z} \left(r v_\phi\right) &= &0. \nonumber\\ \label{eq:AxiSymmetricGammaBis}
\end{eqnarray}

 The axi-symmetric approximation makes possible to perform a first integration of $  \omega_\phi \approx  -\frac{1}{r} \frac{\partial^2 \psi }{\partial z^2} $ leading to the set of local partial differential equations (\ref{eq:AxiSymmetricOmegaBis}) and (\ref{eq:AxiSymmetricGammaBis}) for $\psi$ and $v_\phi$. Nevertheless, it must be emphasized that this model  is ``less non-local'' because, it provides a direct dynamics for $ \frac{\partial \psi }{\partial z }$, however  $ \frac{\partial \psi }{\partial r }$ is also required in equation (\ref{eq:AxiSymmetricOmegaBis}) for a complete specification of the dynamics.
 It should be remarked, that the approximation considered here   (\ref{eq:KasnerCriteria})  for the present model (\ref{eq:AxiSymmetricOmegaBis}) and (\ref{eq:AxiSymmetricGammaBis}) is the opposite of Barkley's work \cite{Barkley2020}.

\subsection{Time dependent reduced model}
\label{Sec:FinalModel}
 
By defining the following new variables~:
\begin{eqnarray} q= r^2 ,\quad  q  J=  \frac{\partial \psi }{\partial z }  ,\quad     z  K = \frac{\partial \psi }{\partial q} \quad \& \quad v_\phi =\sqrt{q}  w ,\nonumber\\ \label{eq:NewVariables}
\end{eqnarray}
and by introducing this  change of variable into equations (\ref{eq:AxiSymmetricOmegaBis}) and (\ref{eq:AxiSymmetricGammaBis}), one gets the following coupled system of partial differential equations~:  
\begin{eqnarray}
\frac{\partial J }{\partial t} -2 q J  \frac{\partial J }{ \partial q}   + 2 z K \frac{\partial J }{\partial z} & = &  {J^2 } -  w^2  , \label{eq:AxiSymmetricJ}\\
 \frac{\partial w }{\partial t} -2 q J  \frac{\partial  w   }{\partial q}  +  2 zK \frac{\partial w  }{\partial z}  &= & 2 J w, \label{eq:AxiSymmetricw}\\
  \frac{\partial \left( q J \right)}{\partial q }  &= &   \frac{\partial \left( zK \right)}{\partial z} . \label{eq:AxiSymmetricK}
\end{eqnarray}

Equation (\ref{eq:AxiSymmetricK}) follows from the condition $$\frac{\partial^2\psi }{\partial r\partial z } = \frac{\partial^2\psi }{\partial z\partial r } ,$$ and, more important, it traces back the non-local aspect of the original Euler equation already discussed.

Equations (\ref{eq:AxiSymmetricJ}) and (\ref{eq:AxiSymmetricw}) are complemented with the initial conditions:
\begin{eqnarray}J(q,z, t=0)= J_0(q,z), \, & & \, w(q,z, t=0)= w_0(q,z) .
\label{eq:InitialCondition}
\end{eqnarray}
Lastly, the boundary conditions read 
\begin{eqnarray}
J(q=0,z, t)= {\mathcal J}_0<\infty , \quad & & \quad w(q=0,z, t)= {\mathcal W}_0 <\infty ,\nonumber\\
\lim_{q\to \infty} J(q,z, t) \to 0  , \quad & & \quad  \lim_{q\to \infty} w(q,z, t)\to 0,\nonumber\\
\lim_{z\to \pm \infty} J(q,z, t) \to 0  , \quad & & \quad \lim_{z\to \pm \infty} w(q,z, t)\to 0,
\label{eq:BoundaryCondition}
\end{eqnarray}
and must be consistent  with a  finite energy solution. 

It should be remarked, that the question of existence or not of a solution exhibiting a finite-time singularity does not depend on the knowledge of $K(q,z, t)$. The vertical velocity and the field $K(q,z, t)$ are both estimated later in Sec. \ref{Sec:VerticalSpeed}.

\section{Singularity behavior in the axi-symmetric simplified model}\label{Sec:Singular}
\subsection{Solution of equations  (\ref{eq:AxiSymmetricJ}) and (\ref{eq:AxiSymmetricw})  by Riemann's characteristic method.} 
 It can be noticed, that the set of partial differential equations (\ref{eq:AxiSymmetricJ},\ref{eq:AxiSymmetricw}) together with the initial conditions (\ref{eq:InitialCondition}) are written in an Eulerian fashion in which the coordinates $(q,z)$ are fixed in time. 
In the following, the hyperbolic system (\ref{eq:AxiSymmetricJ},\ref{eq:AxiSymmetricw})  is solved by the method of characteristics. Basically it consists by passing from an Eulerian description to a Lagrangian one which reduces (\ref{eq:AxiSymmetricJ}) and (\ref{eq:AxiSymmetricw})  into a system of four ordinary differential equations (o.d.e.)~:
\begin{eqnarray}
    \frac{d q  }{d t} & =&    -2 q J  
    , \label{eq:Hamiltonq} \\
  \frac{d z  }{d t} & =&  2 z K, \label{eq:Hamiltonz} \\
  \frac{d J }{d t}    & = &  {J^2 } -  w^2  , \label{eq:J}\\
 \frac{d w  }{d t}  &= & 2 J w. \label{eq:w}
      \end{eqnarray}

This o.d.e. system is complemented with the continuous set of initial conditions:
    \begin{eqnarray}
q(q_0,z_0, t=0)   & = &q_0    , \label{eq:qt=0}\\
z(q_0,z_0, t=0)   & = &  z_0 , \label{eq:zt=0}\\
J(q_0,z_0, t=0)   & = & J_0(q_0,z_0)  , \label{eq:Jt=0}\\
w(q_0,z_0, t=0)   & = & w_0(q_0,z_0) .\label{eq:wt=0}
    \end{eqnarray} 
The initial conditions (\ref{eq:qt=0}-\ref{eq:zt=0}) are such that original Eulerian initial condition (\ref{eq:InitialCondition}) is perfectly parametrized by $(q_0,z_0)$. 
 
As already mentioned, condition (\ref{eq:AxiSymmetricK}) is only required for the integration of $z(t)$ by using equation (\ref{eq:Hamiltonz}). Otherwise, equations (\ref{eq:Hamiltonq},\ref{eq:J},\ref{eq:w}) are independent of  (\ref{eq:Hamiltonz}), in this sense the variable $z(t)$ just follows, neither in an easy or direct way, the dynamics of other variables $q(t),$ $J(t)$ and $w(t)$.

\subsection{The Eulerian-Lagrangian passage of the boundary conditions.}\label{Sec:LagrangianBC}

The boundary conditions for the original problem (\ref{eq:BoundaryCondition}) must be consistent with the Lagrangian solution by the use of the dynamical system  (\ref{eq:Hamiltonq}-\ref{eq:w}). In particular, the boundary conditions at the axe of rotation are such that both velocities, $v_r=v_\phi=0$ at $r=0$. The above implies~: 

 \begin{eqnarray}
 \lim_{r\to 0} v_r  &=&  - \lim_{q\to 0} \sqrt{q} \, J(q,z,t)  =0 ,\nonumber \\
  \lim_{r\to 0} v_\phi &=&  \lim_{q\to 0}  \sqrt{q} \, w(q,z,t)   =0  .\label{eq:BCaxis}
   \end{eqnarray}
Therefore, it is assumed that both, $J(0,z_0, t) $ and $w(0,z_0, t)$, remain bounded, a result shown in the following section. Therefore, the boundary conditions (\ref{eq:BCaxis}) at the axis of rotation are fully satisfied.
Indeed, the axis is  characterized by an initial condition for the dynamical system (\ref{eq:Hamiltonq}-\ref{eq:w}), {\it i.e.}
$q_0=0$ for $z_0 \in (-\infty,\infty)$. 
Finally, the axis is well characterized by:
\begin{eqnarray}
q(0,z_0, t=0)   & = &0   , \label{eq:Axisqt=0}\\
z(0,z_0, t=0)   & = &  z_0 , \label{eq:Axiszt=0}\\
J(0,z_0, t=0)   & = & J_0(0,z_0) < \infty  , \label{eq:AxisJt=0}\\
w(0,z_0, t=0)   & = & w_0(0,z_0)<\infty .\label{eq:Axiswt=0}
    \end{eqnarray} 
  More important, the evolution, as given by equation (\ref{eq:Hamiltonq}), preserves the axis, $q=0$, since 
    \begin{eqnarray}
q(0,z_0, t)   & = &0   , \quad \forall t\geq 0 \quad \&\quad  -\infty< z_0< \infty  . \label{eq:Axisq=0}    \end{eqnarray} 
In the terminology of the theory of dynamical systems, the axis $r=0$ ($q=0$) is called a fixed point of system (\ref{eq:Hamiltonq},\ref{eq:J},\ref{eq:w}) (See Appendix \ref{Sec:Mapping}).

\subsection{ Exact solution of the dynamical system (\ref{eq:Hamiltonq}), (\ref{eq:J}) and (\ref{eq:w}). }
\label{Sec:ExactSolutionJw}

Equations (\ref{eq:J}) and  (\ref{eq:w}) can be directly integrated by setting $Z= J + i w$, which leads to~: $$\frac{d Z }{d t}   =Z^2 ,$$ whose  general solution is~:
   \begin{eqnarray}
Z(q_0,z_0,t) = \frac{1}{C(q_0,z_0)  -t},\label{eq:Z(t)}
    \end{eqnarray}
and, where $C(q_0,z_0) $ is a complex number related to the initial condition by:
$$C(q_0,z_0)  =   \frac {1}{J_0(q_0,z_0)+ i  w_0(q_0,z_0)}  
.$$
In what follows, in order to simplify notations, we set $C(q_0,z_0) \equiv \tau(q_0,z_0) - i \Delta (q_0,z_0) $ an expression which contains all the required information regarding the initial condition.
By splitting into real and imaginary parts in (\ref{eq:Z(t)}) one gets:
   \begin{eqnarray}
  J(q_0,z_0,t) &=&
  \frac{ \tau(q_0,z_0)  -t}{( \tau(q_0,z_0)  -t)^2 +  \Delta(q_0,z_0) ^2 }  ,\label{eq:J(t)}\\
 w(q_0,z_0,t) &=&
  \frac{ \Delta(q_0,z_0) }{( \tau(q_0,z_0)  -t)^2 + \Delta(q_0,z_0) ^2 }  .\label{eq:w(t)}
    \end{eqnarray}

Next, we proceed by  integrating (\ref{eq:Hamiltonq}), obtaining:
 \begin{eqnarray}
  q(q_0,z_0,t) &= & q(q_0,z_0,0) e^{-2 \int_{0}^tJ(t') dt'} \nonumber\\
  &=&  q_0  \frac{ ( \tau(q_0,z_0)-t )^2 +  \Delta(q_0,z_0) ^2 }{\tau(q_0,z_0)^2+  \Delta(q_0,z_0) ^2 } .\label{eq:q(t)}
   \end{eqnarray}

 In what follows, we explore the consequences in the behavior of the exact solutions as given by equations (\ref{eq:J(t)}), (\ref{eq:w(t)}) and (\ref{eq:q(t)}).

\subsection{  On the appearance of a multivalued velocity flow. }\label{Sec:Multivalued}

The solution of the dynamical system (\ref{eq:Hamiltonq}-\ref{eq:w}) provides well defined expressions for $J(q_0,z_0,t)$, $w(q_0,z_0,t)$  and $q(q_0,z_0,t)$, as functions of  parameters $q_0$ and $z_0$ (eqns. (\ref{eq:Axisqt=0}) and (\ref{eq:Axiszt=0})) and of the time $t$. However,  the mapping from the Lagrangian to an Eulerian description, allows to show that $J(q,z,t)$  and $w(q,z,t)$ may not necessarily be a single valued function of $q$ predicting a Riemann's-like singularity at some time $t_c>0$.

 Usually, this kind of mechanism is generic and needs only that the coordinate $q(q_0,z_0,t) $, given by (\ref{eq:q(t)}), becomes a saddle in a point $(q_0^{(c)},z_0^{(c)}) $ at some time $t_c$. Qualitatively, following equation (\ref{eq:q(t)}), initially ($t=0$) the coordinate  $q(q_0,z_0,t) $ represents an inclined plane with unit slope along the $q_0$ direction, however, as time goes this plane is deformed making possible the appearance of a saddle point.
 
The conditions for the existence of such a saddle, read:
\begin{widetext}
\begin{eqnarray} \left. \frac{\partial  q  }{\partial q_0}\right|_{q_0^{(c)},z_0^{(c)},t_c}  =  0 , \quad  \left. \frac{\partial  q  }{\partial z_0}\right|_{q_0^{(c)},z_0^{(c)},t_c}  =  0 , \quad \&  \quad 
\left|  \begin{array}{cc}
\frac{\partial^2  q  }{\partial q_0^2} & \frac{\partial^2  q  }{\partial q_0 \partial z_0}\\
\frac{\partial^2  q  }{\partial z_0\partial q_0} & \frac{\partial^2  q  }{ \partial z_0^2}
\end{array}
  \right|_{q_0^{(c)},z_0^{(c)},t_c}   = 0. \label{eq:SaddleCondition} \end{eqnarray}
\end{widetext}
The saddle condition (\ref{eq:SaddleCondition}) determines a critical time, $t_c$, for which the multivalued behavior manifests for a first time as well as its location, $(q_0^{(c)},z_0^{(c)})$, in terms of the initial conditions.  In the Eulerian description the parametric representation of $J(t)$ vs. $q(t)$ shows a singular behavior for $t= t_c$  at a circular rim of finite radius $r_c= \sqrt{ q_c}$. Through, equations (\ref{eq:J(t)}), (\ref{eq:w(t)}), it is noticed that both $J(t_c)$ and $w(t_c)$ are finite but their derivatives $\partial J/\partial q$ or $\partial w/\partial q$  diverge. For $t\gtrsim t_c$, the functions $J(t)$  and $w(t)$ becomes multivalued functions of $q$ in a  region near $q \approx q_c$.

To illustrate this transition in a simpler way, let's consider the saddle on the $q_0$-direction such that the saddle condition (\ref{eq:SaddleCondition}) arises at $q_0^{(c)}$ finite and $z_0^{(c)} =0$ (Because of translational invariance along the $z$ axis it is possible to set  $z_0^{(c)} =0$).
From the catastrophe theory, it can be seen that near the transition point the coordinate $q$ behaves locally ($q_0 \approx q_0^{(c)}$) as:

\begin{widetext}
\begin{eqnarray} 
  q(q_0,z_0,t) \approx  q_c   +  A  (t_c-t)  \left(q_0 -q_0^{(c)}\right) + \frac{1}{3}  B \left(q_0 -q_0^{(c)}\right) ^3 +\frac{1}{2}C  z_0^2 + \dots, \label{eq:Cathastrophe}
  \end{eqnarray}
\end{widetext}
  where $A$, $B$ and $C$ are constants and $q_c=q\left(q_0^{(c)},t_c\right) $ is the value of $q$, at the critical point imposed by the condition (\ref{eq:SaddleCondition}).

After (\ref{eq:Cathastrophe}), the basic scaling suggests:
 \begin{widetext}
  \begin{eqnarray} 
q-  q_c \sim (t_c-t)^{3/2} ,\quad q_0 -q_0^{(c)} \sim (t_c-t)^{1/2}, \quad {\rm and} \quad z_0\sim (t_c-t)^{3/4}. \label{eq:CathastroopheScaling}
 \end{eqnarray} 
  \end{widetext}

 Notice that from these scaling laws, the scaling of the $z$ coordinate cannot be accessed. However, the scaling for the flow velocities,  $ v_r (r, z, t) $ and $ v_ \phi (r, z, t) $,
follows from formulae:
\begin{eqnarray} v_r &=& - r J =  -\sqrt{q(q_0,z_0,t) }  J(q_0,z_0,t) , 
  \label{eq:vr}\\
v_\phi  &=&  r w = \sqrt{q(q_0,z_0,t) }  w(q_0,z_0,t) . 
\label{eq:vphi}  \end{eqnarray}

 At the singularity point, $t=t_c$ and $q=q_c$, the value of $J(t_c)$ and $w(t_c)$ are finite, whence both, $v_r(t_c)$ and $v_\phi(t_c) $, are also finite. However, their radial derivatives becomes singular. Indeed, the singular behavior comes from expression (\ref{eq:Cathastrophe}), which leads to
   \begin{eqnarray} 
\frac{\partial v_r}{\partial r} &=& 2 \sqrt{q} \frac{\partial v_r}{\partial q} =  - J (q) - 2 q  \frac{\partial J}{\partial q} .
\label{eq:CathastroopheScalingForSpeed}
 \end{eqnarray} 
 
 Because the singularity arises on a rim at finite $q\approx q_c$ and $J(q_c) $ is also finite, it means that the first term in the previous equation is finite. By writing the second term as 
$$ \frac{\partial J}{\partial q}=  \left( { \frac{\partial J}{\partial q_0^{(c)}}} \right) / \left({\frac{\partial q}{\partial q_0^{(c)}} }\right) ,$$ one notices that after (\ref{eq:Cathastrophe}) or (\ref{eq:CathastroopheScaling}) the denominator vanishes as $(t_c-t)$, while the numerator is regular. Therefore, the radial velocity gradient on the local plane on which the swirl velocity, $v_\varphi$ vanishes, 
scales as

 \begin{eqnarray} 
\frac{\partial v_r}{\partial r} &\approx & -  \frac{\gamma}{(t_c-t)} ,  \label{eq:Vrvst}
 \end{eqnarray} 
where the constant 
 \begin{eqnarray} 
 \gamma = \frac{2 q_c}{A} \left. \frac{\partial J}{\partial q_0}\right|_{q_0^{c}} ,\label{eq:gamma}
 \end{eqnarray} 
 depends on the parameters at the critical point, and more importantly, it depends on the initial conditions of the flow, through $J_0(q_0,z_0)$ and  $w_0(q_0,z_0)$.
Consequently, by the condition of incompressibility one also expects $\frac{\partial v_z}{\partial z}\sim \frac{1}{(t_c-t)} $.  Although, other components of the full stretching tensor, like $\partial_r v_z$, $\partial_z v_r$, and vorticity components cannot be determined without the explicit knowledge of $z$ or $v_z$,  the energy dissipation rate:
\begin{widetext}
 \begin{eqnarray} 
  \partial_k v_i \partial_k v_i  &=&    \frac{v_r^2+v_\varphi^2}{r^2}  +  \left( \frac{\partial v_r} {\partial r} \right)^2 +\left( \frac{\partial v_r} {\partial z} \right)^2 + \left( \frac{\partial v_\varphi} {\partial r} \right)^2+ \left( \frac{\partial v_\varphi} {\partial z} \right)^2 + \left( \frac{\partial v_z} {\partial r} \right)^2 + \left( \frac{\partial v_z} {\partial z} \right)^2,
  \label{eq:DissipationRate}
 \end{eqnarray} 
   \end{widetext}
   diverges at least as $1/(t_c-t)^2$. 

Summarizing, in this section it is shown that the general evolution of the velocity field gives rise to a Riemann-like mechanism for which $\frac{\partial v_r}{\partial r}$, as well as, $\frac{\partial v_\phi}{\partial r}$ both diverge at a circular rim. Others, velocity gradient may also diverge as discussed at the end of the paper. Moreover, in the case of zero swirl velocity, $v_\phi=0$, a singularity is also expected, but only $\frac{\partial v_r}{\partial r}$ and $\frac{\partial v_z}{\partial z}$ diverge in finite-time, while other velocity gradients or vorticity components remain bounded (See Section \ref{Sec:ZeroSwirl}).

Finally, notices that as it occurs for compressible fluids, this transition from a single to a multi-valued flow could possibly be regularized by viscosity. We shall investigate this issue in a future publication.

\subsection{The vertical velocity field.}\label{Sec:VerticalSpeed}
Up to this stage, through eqns. (\ref{eq:Hamiltonq}-\ref{eq:w}) the variables $J(q_0,z_0,t)$, $w(q_0,z_0,t)$  and $q(q_0,z_0,t)$ are solved explicitly, however the function $z(q_0,z_0,t)$, requires the knowledge of $K(q_0,z_0,t)$, which hides the non-local structure of the incomprensible Euler equations. However, the velocity field may be estimated by making use of the magnetostatic analogy  (${\bm \nabla} \cdot {\bm v}=0$ and ${\bm \nabla} \times {\bm v}={\bm \omega}$) through the Biot-Savart law (\ref{eq:BiotSavart}) for an axi-symmetric configuration (This calculation is inspired by a note on Elgindi's approximation by T. Tao \cite{Tao2019})~:
\begin{widetext}
\begin{eqnarray}
\psi(r,z,t) &= &\frac{r}{4 \pi} \int \frac{\cos \phi' } { \left( r'^2+ r^2 - 2 r' r \cos\phi'+ ( z-z')^2\right)^{1/2} } \, \omega_\phi(r',z',t') \, r' dr' d\phi' d z' .\nonumber
 \end{eqnarray}

Assuming 
$  \omega_\phi (r',z',t)=  -r' \frac{\partial J }{\partial z'} $ for consistency, and 
after an integration by parts which requires a decaying dependence of $J(r',z',t) $ as $z'\to\pm\infty$, it is obtained: 
\begin{eqnarray}
\psi(r,z,t) &= &\frac{1}{4 \pi} \int \frac{ ( z-z') r {r'} \cos \phi' } { \left( r'^2+ r^2 - 2 r' r \cos\phi'+ ( z-z')^2\right)^{3/2} } J(r',z',t)  \, r' dr' d\phi' d z' . \nonumber   
 \end{eqnarray}

Computing the vertical  velocity via (\ref{eq:DefStream}) gives the following expression for the vertical velocity:
\begin{eqnarray}
v_z &=&\frac{1}{4\pi r} \int \frac{ ( z-z') {r'} \cos \phi' \, J(r',z',t) } { \left( r'^2+ r^2 - 2 r' r \cos\phi'+ ( z-z')^2\right)^{3/2} }  \, r' dr' d\phi' d z'  - \frac{3}{4\pi } \int \frac{ ( z-z') {r'} \cos \phi' (r-{r'} \cos \phi' ) \, J(r',z',t) } { \left( r'^2+ r^2 - 2 r' r \cos\phi'+ ( z-z')^2\right)^{5/2} }  \, r' dr' d\phi' d z' .
\nonumber\\
\label{eq:BSvzFull}       \end{eqnarray}
\end{widetext}

It is noticed that because $|J(r',z',t)|$ is bounded, then the inner behavior, {\it i.e.} for $ r' < r$ and $ |z'| < |z|$, the  vertical velocity (and the radial velocity) scales as a length. This property known from electrostatic: the electric field generated by a uniform density charge is linear in distances, was used by Elgindi to approximate the Biot-Savart integral. In short, the integrals in (\ref{eq:BSvzFull}) are approximated taking $r'\gg r$ via a multipolar expansion. After a direct calculation it is obtained:
\begin{eqnarray}
v_z &\approx &\frac{3z}{2} \int_ {r' > r}  \left(  \int_ {|z'| > |z|}   \left( z-z' \right) J(r',z',t) d z' \right)  \, \frac{ dr' }{ {r'}^2 } .
\label{eq:BSvz}       \end{eqnarray}

Similarly, the leading order for the radial velocity gives: 
\begin{eqnarray}
v_r&\approx &-\frac{3r}{4} \int_ {r' > r}   \left(  \int_ {|z'| > |z|}   J(r',z',t) dz'\right)\, \frac{ dr' }{ {r'}^2 }  .
\label{eq:BSvr}       \end{eqnarray}

In particular, in the special case of an odd-symmetry (See  the next Sec. \ref{Secc:QualitativeFlow}), the field $ J(r',z',t) $ is an even function respect to the plane $z'=0$, then, the integration of $ \int_ {|z'| > |z|} {z'}  J(r',z',t)  d z'$ vanishes in (\ref{eq:BSvz}).  By comparing both expressions, (\ref{eq:BSvz}--\ref{eq:BSvr}), 
up to leading order in $r$, it is found that
$$v_z \approx - 2 z \frac{\partial v_r} { \partial r } \approx  \frac{2\gamma z}{(t_c-t)}.$$
Therefore, the function $K$ must scale also as:  $K \sim  \frac{1}{(t_c-t)}$,
whence $$\frac{dz}{dt} = 2 K z =  -  \frac{\beta_3}{(t_c-t)} z,$$ thus
$$ z(t) \sim (t_c-t)^{\beta_3}.$$
The parameter $\beta_3=-2 \gamma$ does not seem to be universal, in the sense of phase transitions or catastrophe theory, but it depends explicitly on the parameter $\gamma $ defined in (\ref{eq:Vrvst}) and also on the more accurate value of the integrals (\ref{eq:BSvr}) and (\ref{eq:BSvz}). 

The main lesson of this section, is that the hypothesis of an anisotropic exponent is plausible, the value and sign of $\beta_3$ depend essentially on the initial value for the velocity field. 

\subsection{Qualitative singular flow.}\label{Secc:QualitativeFlow}

To fix ideas, let us consider a situation in which one considers an anti-symmetric initial flow \cite{Luo2014,Barkley2020}. This flow is invariant under $ z \to -z $ ($ z_0 \to -z_0 $), $ v_ \phi \to -v_ \phi $, $ v_z \to -v_z $ and $ v_r \to v_r $. The flow corresponds to an initial condition that makes the swirl velocity to vanish in the line $ z_0 = 0, \, \forall q_0 \geq 0 $. A simple choice would be $ \Delta (q_0, 0) =0$, and  $\tau (q_0, 0)   = \tau_0(q_0) $, where $\tau_0(q_0)$  is a function which has a minimum at $q_0^*$ and $\tau(q^*_0)=t_*>0$.
 The advantage in choosing this kind of up-down symmetric flow is that the velocity flow is exactly computed on the plane $z=0$. The velocity components are: $v_r=  -\sqrt{q} \, J(q,z=0,t)$, $v_\phi =0$, and $v_z=0$ at $z=0$. Therefore the scaling (\ref{eq:Vrvst}) becomes exact at the plane of symmetry $z=0$.

 To illustrate the multi-valued behavior,  the initial condition can be modeled by setting~:
  \begin{eqnarray} 
 \tau (q_0, z_0)   & = &\tau_0(q_0) = t_* + a (q_0-q_*)^2  , \label{eq:InitTau}\\
 \Delta (q_0, z_0)  &=& b z_0. \label{eq:InitDelta}
 \end{eqnarray}
 Although, the above initial condition exhibits an infinite energy flow, the Ansatz (\ref{eq:InitTau}) and (\ref{eq:InitDelta}) characterizes the relevant features that the initial flow must possess, that is $ \Delta (q_0, z_0)  $ vanishes and $ \tau (q_0, z_0) $ has a minimum both in the same line. 
 
 In the current case, on the plane $z=0$, the solutions for $J$ and $q$ read ($w$ vanishes exactly)
\begin{eqnarray}J(t)   & = & 
  \frac{1}{ \tau_0(q_0)  -t} , \label{eq:JDelta0}\\
   q(t)&=&  q_0 \left( 1- \frac{ t }{\tau_0(q_0)}\right)^2. \label{eq:qDelta0}
  \end{eqnarray}

\begin{figure}[h]
\centerline{ (a) \includegraphics[width=0.45\columnwidth]{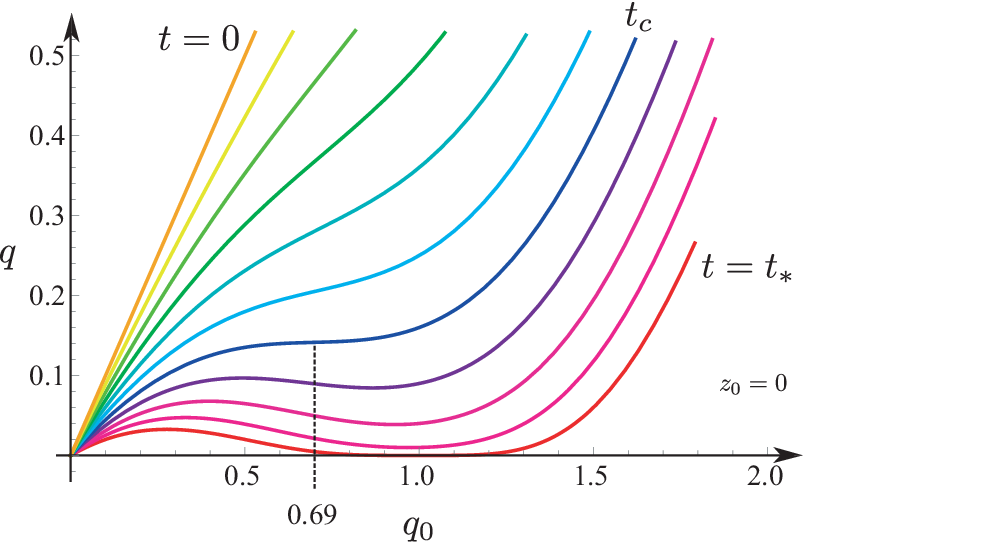}\, (b) \includegraphics[width=0.45\columnwidth]{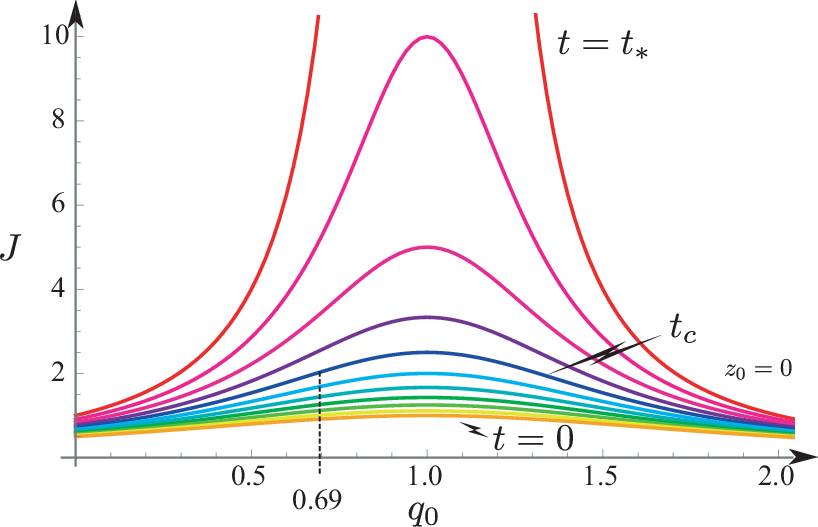}}
\caption{(a) Plot of the radial coordinate $q$ vs $q_0$. The saddle-condition (\ref{eq:SaddleCondition}) is sketched in the plot. The critical saddle-transition appears for $q_0 ^c \approx 0.69$ and for a time $t_c \approx 0.61$. (b) Plot of the function $J(q_0)$ (\ref{eq:JDelta0}) as a function of  $q_0$. The  plots are done using the initial flow represented by (\ref{eq:InitTau}) and (\ref{eq:InitDelta}), with the parameters  $t_*=q_*=a=b=1$. In the above figures the curves are plotted  at $z_0=0$ and for times spanning into in $0\leq t\leq t_*$.} 
\label{Fig:MultivaluedTransition}
\end{figure}

Therefore, after equation (\ref{eq:q(t)}) one notices that as $t$ increases the function $q( q_0, z_0,t)$ changes its monotonicity: it switches from a monotonic increasing function for $t<t_c$ to a non-monotonic function with at least one maximum and a minimum  for $t>t_c$.
 FIG. \ref{Fig:MultivaluedTransition}-(a) shows the change of concavity in the radial coordinate $q$ vs. $q_0$. Additionally, FIG. \ref{Fig:MultivaluedTransition}-(b) shows the regular behavior of the the function $J(q_0,t)$ at the critical point.

In that situation, one notices that a given $q$ may arise from three different values of $q_0$, thus  $J(q)$ takes three different values. 
In this way, the function $J$ becomes multivalued, this transition from a single-valued to a multi-valued function represents a singularity.

\begin{figure}[h]
\centerline{(a) \includegraphics[width=0.45\columnwidth]{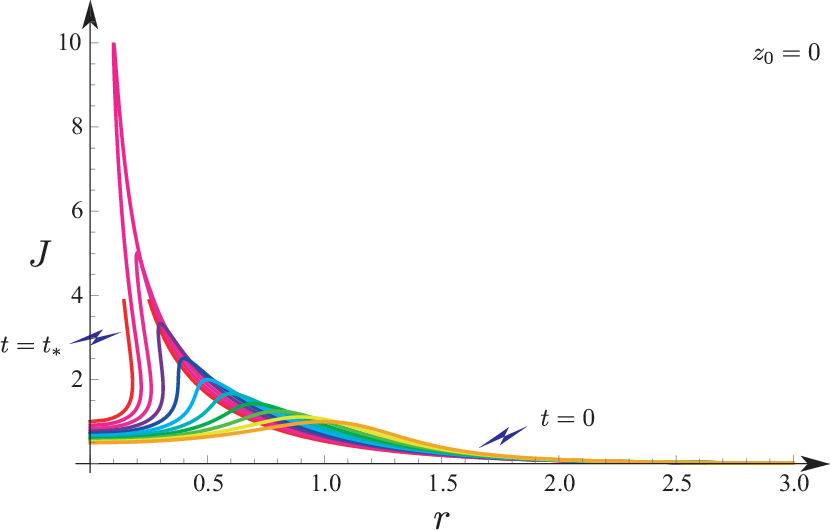} \, (b) \includegraphics[width=0.45\columnwidth]{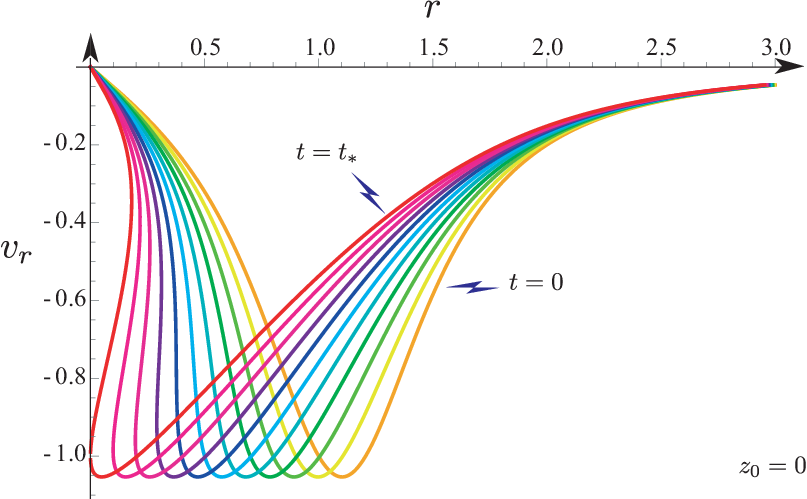} }
\caption{(a) Parametric plot (with $q_0$ as the parametrization) of the $J(q_0)$ (\ref{eq:JDelta0}) as a function of the distance from the axis of rotation $ r(q_0) $ (\ref{eq:qDelta0}). (b) Parametric plot of the function $v_r(q_0)$ as a function of the distance from the axis of rotation $ r(q_0) $.  
The plots are done using the same representation as in FIG. \ref{Fig:MultivaluedTransition}.  The swirl velocity is zero for a given initial condition such that $\Delta=0$. }
\label{Fig:Multivalued}
\end{figure}

 FIG. \ref{Fig:Multivalued}-(a) displays parametrically the function, $J$, as function of the radial distance, $r$; and, FIG. \ref{Fig:Multivalued}-(b) shows the radial velocity, $v_r$, vs. $r$. Both plots show curves for various times  such that $t\leq t_*$. 
One notices that initially both $J$ and $v_r$ in FIGs. \ref{Fig:Multivalued}-(a) \& (b)  are single valued functions, but they become multivalued as soon as $t>t_c$. 

\begin{widetext}

\begin{figure}[ht]
\centerline{(a1) \includegraphics[width=0.3\columnwidth]{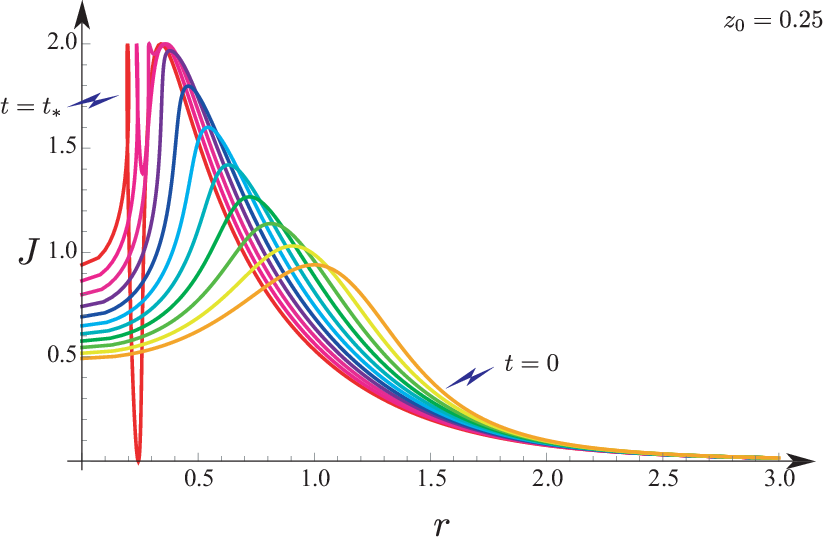}\, (b1) \includegraphics[width=0.3\columnwidth]{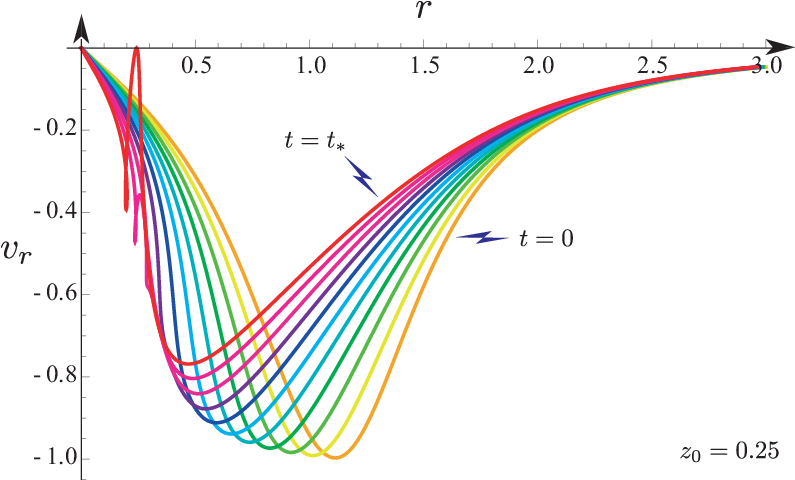} \, (c1)  \includegraphics[width=0.3\columnwidth]{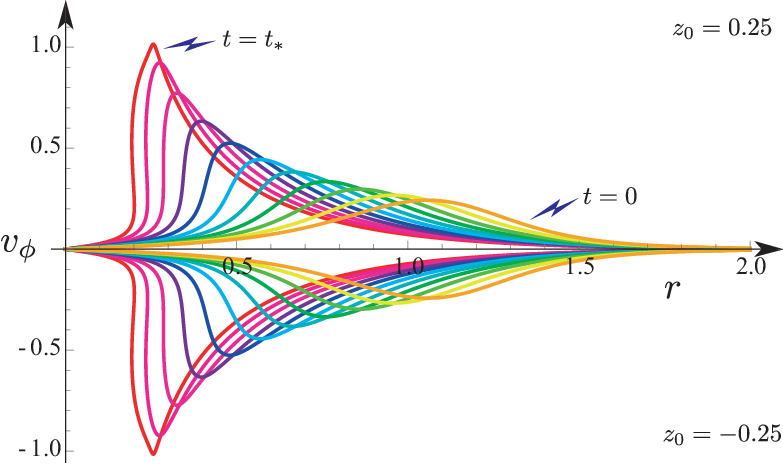} }
\centerline{(a2) \includegraphics[width=0.3\columnwidth]{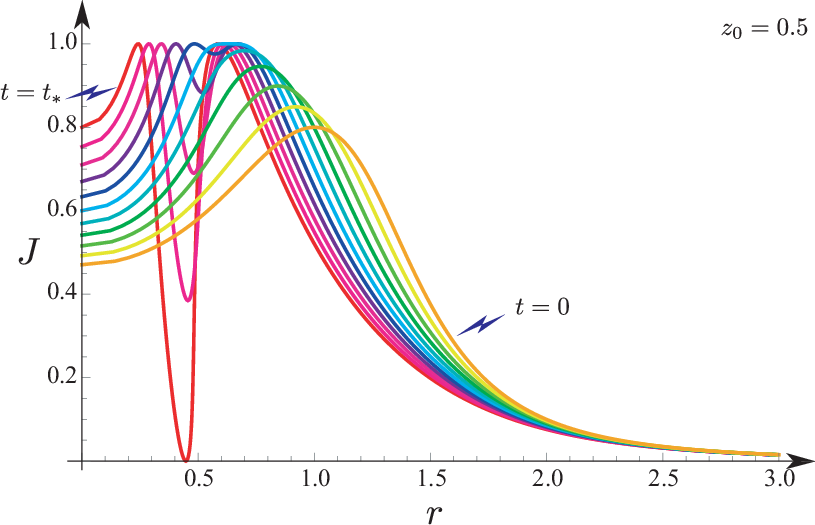}\, (b2) \includegraphics[width=0.3\columnwidth]{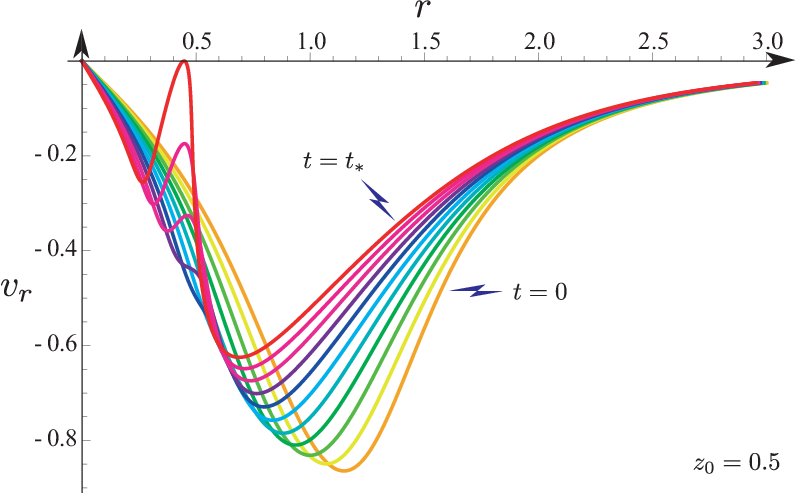} \, (c2)  \includegraphics[width=0.3\columnwidth]{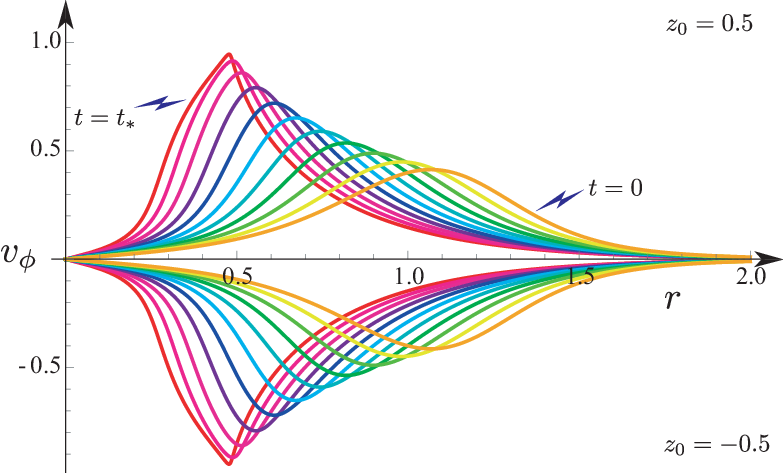} }
\caption{Parametric plot of the functions: (a1) \& (a2)  $J$;  (b1) \& (b2)  $v_r$, and,  (c1) \& (c2)   $ v_ \phi $ vs. the distance from the axis of rotation $ r $, for different times varying between $ t = 0 $ to $ t = t _ * $.  The plots are done using the same parameters as in FIG. \ref{Fig:MultivaluedTransition}. The top the row 1 corresponds to
$z_0=1/4$, while, the second row corresponds to  $z_0=1/2$. For the case $z_0=1/2$ the quantities $J$, $v_r$ and $v_\phi$ are single valued functions, for all values of $r$.} \label{fig:Multivalued2}
\end{figure}
 \end{widetext}

In the following we focus on the qualitative behavior of the flow velocities:   $ v_r (r, z, t) $ and $ v_ \phi (r, z, t) $ as given by (\ref{eq:vr}) and (\ref{eq:vphi}) for the fluid velocity off the $z_0=0$ plane. FIG. \ref{fig:Multivalued2} shows $J$, $v_r$ and $v_\phi$ as a function of $r$ for different times as well as different initial planes $z_0=1/4$ and $z_0=1/2.$ One notices that all quantities remain multivalued into the neighborhood of $z_0=0$, however these may become single valued as $z_0$ increases.

As has been already shown,  both $v_r$ (\ref{eq:vr}) and, $v_\phi$ (\ref{eq:vphi}) vanish at the axis of rotation, $r=0$.  Moreover, one notices the existence of a shear structure of the flow defined by the plane in which $\Delta(q_0,z_0) $ vanishes. 
For points such that initially 
$ \Delta (q_0,z_0)\approx0,$ the radial and the swirl velocity read
  \begin{eqnarray}
 v_r  &\approx& - 
  \frac{( t_*  -t)  }{ \sqrt{ ( t_*  -t)^2 + \Delta(q_0,z_0) ^2} }  , \nonumber \\
 v_\phi &\approx &\frac{ \Delta(q_0,z_0) }{ \sqrt{ ( t_*  -t)^2 + \Delta(q_0,z_0) ^2} } . \nonumber 
   \end{eqnarray}

\begin{figure}[h] \includegraphics[width=0.7\columnwidth]{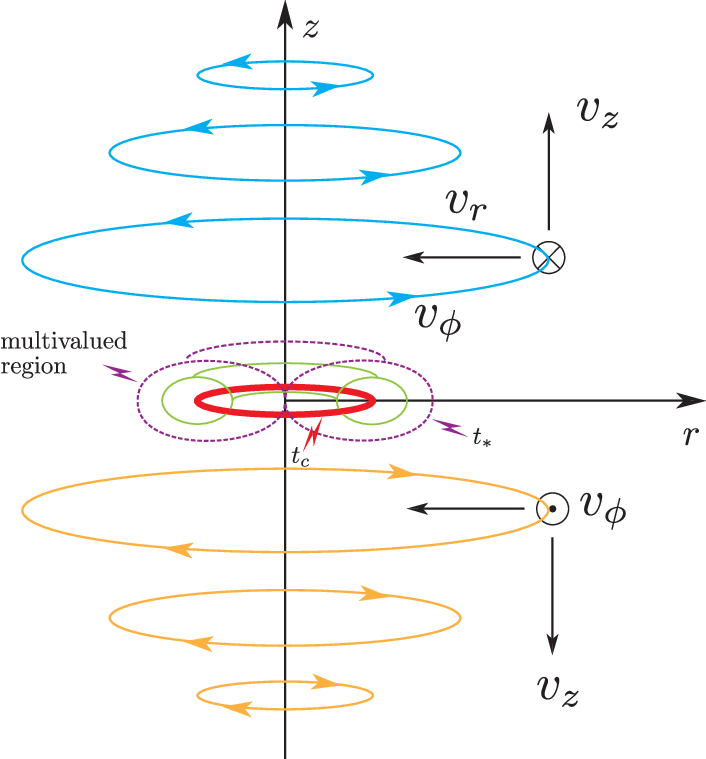}
\caption{ Qualitative sketch of the singular flow. The radial velocity, $v_r$, is directed into the axis, the vertical velocity, $v_z$, shows an outflow coming from the $z=0$ plane, the azimuthal velocity $v_\phi$ is drawn into the plane by a symbol $\otimes$ for $z>0$ and out of the plane the by the symbol $\odot$ for $z<0$ showing a tangential discontinuity. The curves represent the qualitative behavior of streamlines attained by the swirl flow.
The swirl velocity $ v_ \phi $ is directed downwards the plane for $ z> 0 $ and in the opposite upwards direction for $ z <0 $.
The radial velocity, $ v_r $, is directed towards the axis of symmetry. Because of condition $ {\bm \nabla} \cdot{\bm v} = 0 $, the vertical velocity, $ v_z $, must be as shown in the figure. The superposition of the three motions is schematized with streamlines of the swirl flow. The segmented region highlights a spatial zone in which the velocity flow is multivalued. }
\label{fig:f4}
\end{figure}

The radial velocity contracts the flow into the axis, while the swirl flow, $v_\phi$ changes its sign as one crosses the line $\Delta(q_0,z_0)=0$. Lastly, as a result of the divergence free flow condition the $v_z$ component of velocity repels the flow out of the plane defined by $\Delta(q_0,z_0)=0$. Therefore, the flow suffers a tangential discontinuity. A qualitative sketch of the flow  is shown in FIG. \ref{fig:f4} and is based on the exact solutions (\ref{eq:J(t)}) and (\ref{eq:w(t)}) by using the expressions for the initial velocities described by (\ref{eq:InitTau}) and (\ref{eq:InitDelta}).  

The most striking feature of the flow is that at a time $t_c$ the radial velocity gradient $\frac{ \partial v_r }{  \partial  r} $ becomes singular located on the plane $z=0$ at a rim at a finite distance $r_c = \sqrt{q_c}$. Similarly,  $\frac{ \partial v_\phi}{  \partial  r} $  becomes singular in the vicinity of the symmetry plane $z=0$. For $t\gtrsim t_c$, this singular rim grows into a toroidal volume inside which the velocities are multi-valued. As $t\to t_*$ the inner side of the toroidal region reaches the axis of rotation, as sketched in the FIG. \ref{fig:f4}.

\subsection{ On the existence of a finite-time singularity of $J(t)$.}
\label{Sec:ExistenceOfFiniteTimeSingularity}

 In this section we show how the  formal solution (\ref{eq:J(t)}), (\ref{eq:w(t)}) and (\ref{eq:q(t)}) exhibits a divergence in finite-time. Nevertheless, this divergence arises after the formation of multivalued solutions at $t_c$. Although, it appears to be a pure formal singularity, the natural continuation of the solution of (\ref{eq:qt=0}-\ref{eq:zt=0})  into the multivalued domain still presents some interest because it may play a role whenever multi-valued solutions may be regularized by viscosity.  Moreover, this kind of singularity resembles the one found by Elgindi \cite{Elgindi2021annals}, and more, the mathematics has a similarity to the ones which can be found in the works of Constantin, Lax, and Majda \cite{ConstantinLaxMajda1985}, and De Gregorio \cite{DeGregorio1990,DeGregorio1996}. However, the physical mechanisms seem to be different.

 Next, we prove the following statement: {\it Let $(q_0^*,z_0^*)$ be a point such that $\Delta(q_0^*,z_0^*)=0$ and $ \tau(q_0^*,z_0^*)$ is the absolute minimum in the manifold $\Delta(q_0^*,z_0^*)=0$, then,  $  J(q,z,t)$ diverges at time $t_{*}= \tau(q_0^*,z_0^*) >0.$}  This singularity is shown in FIG. \ref{Fig:MultivaluedTransition}-(b) for the $t=t_*$. 

\begin{figure}[h]
(a) \includegraphics[width=0.4\columnwidth]{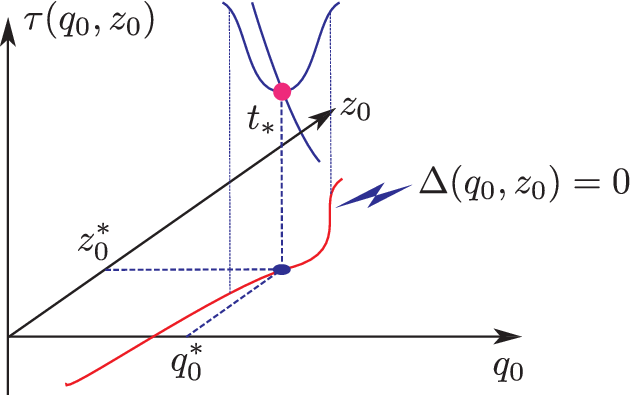} \, (b) \includegraphics[width=0.4\columnwidth]{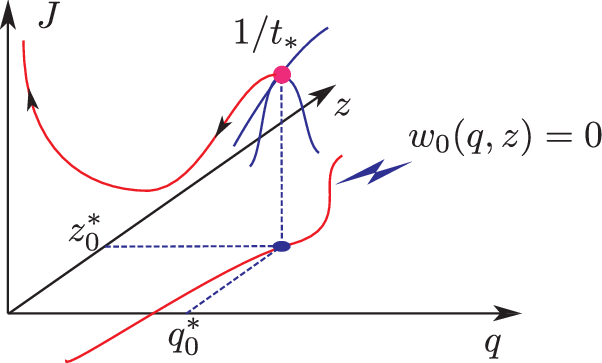}
\caption{(a) A generic initial condition for the o.d.e. system
(\ref{eq:Hamiltonq}-\ref{eq:w}). The parameter  $\Delta(q_0,z_0)$ vanishes on the curve, and the function $\tau(q_0,z_0) $ reaches a minimum along the curve at   $t_*=\tau(q_0^*,z_0^*)$. (b) Sketch of the flow of the dynamical system (\ref{eq:Hamiltonq}--\ref{eq:w}). Initially, $(q,z)$ matches perfectly with $(q_0,z_0)$,  and  $J(q_0^*,z_0^*)\equiv 1/t_*$, then the evolution of $J(t), q(t),z(t)$ drifts away from the original point $(1/t_*,q_0^*,z_0^*)$ to the origin  $q=0$ and $z=0$ (notice that because of translational invariance in $z$ we settled the singularity at $z=0$). Furthermore, at the origin $J\to \infty$.  }
\label{fig:f1}
\end{figure}

The proof of this statement is as follows. 
Consider an initial point 
$(\bar q_0,\bar z_0)$ such that $\Delta(\bar q_0,\bar z_0)=0$,
 then the denominator in (\ref{eq:Z(t)}) is a pure real number, therefore $Z(\bar q_0,\bar z_0,t)$ is real and diverges  as  $$Z (\bar q_0,\bar z_0,t) = \frac{1}{(\tau(\bar q_0,\bar z_0)-t)}, $$ when $t\to  \tau(\bar q_0,\bar z_0)$.  This singular behavior occurs for all points $(\bar q_0,\bar z_0)$  on the curve given by the implicit relation $\Delta(\bar q_0,\bar z_0)=0$. However, the singularity arises first for the minimal value of all possible $\tau(\bar q_0,\bar z_0)$,  {\it i.e.}
 $$ t_* =\min_{\bar q_0,\bar z_0}\{\tau(\bar q_0,\bar z_0)\,|\, \Delta(\bar q_0,\bar z_0)=0\} >0.$$
 Whence  $Z (q_0^*,z_0^*,t) = {1}/(t_* -t)$  as $t\to t_*$ (see FIG. \ref{fig:f1}-a).
 
\medskip

In conclusion, it should be remarked that this later or secondary singularity arises as the radial velocity touches the axis of rotation $r=0$, as seen in FIG. \ref{Fig:Multivalued}-(b), thus at $t_*$ the multivalued region reaches the axis of rotation,  as shown in the Fig \ref{Fig:Multivalued}-(b) for the curve corresponding to $t=t_*$.

\subsection{The case of zero swirl velocity}\label{Sec:ZeroSwirl}
The multivalued nature of solutions of equations  (\ref{eq:Hamiltonq},\ref{eq:J},\ref{eq:w}) appears to be generic and independent of the swirl velocity. Therefore, the singular behavior remains in the case of zero swirl velocity, however in the case of zero swirl velocity, the axial vorticity, driven by 
equation (\ref{eq:AxiSymmetricOmega}), is materially conserved so that
$$\frac{d}{dt} \left( \frac{\omega_\phi}{r}\right)= \frac{d}{dt} \left(- \frac{\partial J}{\partial z}\right)= 0.$$
Moreover, in the case of zero swirl velocity other components of vorticity vanish, {\it i.e.} $\omega_r= \omega_z = 0$. 

As has been shown by Ukhovskii-Yudovich \cite{Ukhovskii1968}, if the initial vorticity is sufficiently smooth (class $C^\infty$), then the axially symmetric flow without swirl is globally regular, excluding any singular behavior of the vorticity in finite-time. However, if the initial condition is differentiable but not sufficiently smooth, then the  global regularity is not known.  Recently, Elgindi \cite{Elgindi2021annals} has shown that an axisymmetric flow without swirl may exhibit a self-similar blow-up in finite-time  if initial condition is sufficiently smooth $C^{1, \alpha}$, which is not excluded by  Ukhovskii-Yudovich Theorem. 

The existence of a multivalued solution exhibiting a singularity of $\frac{\partial v_r}{\partial r} $ and $ \frac{\partial v_z}{\partial z} $ possibly conjectures that, despite initially $J(q,z,t=0)$ is $C^\infty$, as time evolves, higher derivatives of $\frac{\partial J}{\partial z}$ may not exist. This aspect must be regarded carefully in the future.

\section{Discussion and Perspectives}\label{Sec:Discussion}

Under the assumption that  Euler (and Helmholtz) equations generate a spatial anisotropy in time, it is shown that  the axi-symmetric flow with swirl may be approximated by a hyperbolic non-linear system  (eqns. (\ref{eq:AxiSymmetricOmegaBis}) \& (\ref{eq:AxiSymmetricGammaBis})) which is solved using the method of characteristics. It is shown that generically, that solutions of the approximate system of equations  become multivalued in finite-time. Under these special conditions as time reaches a critical time $t_c$, the radial velocity, $v_r$,  and the swirl velocity, $v_\phi$, remain finite but the radial derivatives diverge as: $$\frac{\partial v_r}{\partial r}  \sim \frac{1}{(t_c-t)} \quad {\rm and }\quad \frac{\partial v_\phi}{\partial r}  \sim \frac{1}{(t_c-t)}. $$ 

 A second result is that, if initially, the axial speed, $v_\phi$ vanishes on a line in the $(r,z)$ plane, then the solution of the approximate model will develop a secondary singularity at some later time $t_*> t_c$.   

The complete velocity flow cannot be computed exactly, but it is estimated via the Biot-Savart integral validating a possible hypothesis of the existence of an anisotropic flow. 

Lastly, under the assumption of an initial flow with an up-down symmetry the qualitative velocity flow near the singularity time involves a counter rotative swirl flow which may be decomposed into an inflow to the central axis, together with a counter rotative axial flow, and an outflow from the plane defined by $z=0$. See the scheme in Fig. \ref{fig:f4}.
In that scenario, and assuming that the flow shrinks into the  $z=0$ plane according $z\sim (t_c-t)^{\beta_3}$ (here $\beta_3 $ is unknown), and the radial coordinate shrinks into a rim of finite radius $r_c$ as:  $r-r_c\sim (t_c-t)^{3/2}$ with $ \beta_3>3/2$ (see (\ref{eq:CathastroopheScaling})), then 
accordingly with the observed scaling behaviors, one may try a Leray type approach for equations (\ref{eq:AxiSymmetricOmegaBis}) and (\ref{eq:AxiSymmetricGammaBis}), which is written as 
$$\psi = (t_c-t)^{\beta_3 }  \Psi\left( \frac{r-r_c}{(t_c-t)^{3/2}}, \frac{z}{(t_c-t)^{\beta_3}} \right),$$ with $\beta_3>3/2$,
and
$$v_\phi =  z W\left(\frac{r-r_c}{(t_c-t)^{3/2}}, \frac{z}{(t_c-t)^{\beta_3}} \right),$$
Here the pre-factor $z$ must be included by the odd symmetry of the flow.
Accordingly, to this basic scaling one has:

 \begin{widetext}
  \begin{eqnarray} 
\psi \sim v_\phi \sim(t_c-t)^{\beta_3} ,\quad v_r\sim 1 ,  \quad v_z\sim {(t_c-t)^{\beta_3-3/2} } ,  \quad  \omega_r \sim 1 ,\quad  \omega_\phi \sim \frac{1}{(t_c-t)^{\beta_3} }   \quad {\rm and} \quad \omega_z \sim \frac{1}{(t_c-t)^{3/2} } . \label{eq:FullScaling}
 \end{eqnarray} 
  \end{widetext}
Thus, $v_z\sim (t_c-t)^{\beta_3-3/2} $, so that $v_z\to 0$ as $t\to t_c$. 

Moreover, according to the vorticity scaling: $\lVert \omega \rVert_\infty \sim (t_c-t)^{-\beta_3}$ the Beale, Kato, Majda (BKM) \cite{beale1984} criterion diverges at least as $1/(t_c - t)^{\beta_3-1} $ as $t\to t_c$. In view of the fact that the numerical singularity of Luo and Hou \cite{Luo2014} is spatially isotropic these results cannot be used for the purpose of comparison.  Nevertheless, the results of the simulations indicate that $\beta_3 \approx 3.46$ by employing the BKM criterion, and, $\beta_3\approx 2.91 $ for the scaling for the radial component. Both estimations, are greater than $ 3/2$, as expected.

On the other hand, it is easy to see that the resulting blow-up solution has a finite energy (\ref{eq:Energy}). For $\beta_3>3/2$ the kinetic energy coming from $v_z^2$ does not contribute, therefore the convergence only concerns:  $$v_r^2+v_\phi^2  =   q \left( J^2+w^2 \right) \equiv\frac{q_0 }{\tau(q_0,z_0)^2+  \Delta(q_0,z_0) ^2 }, $$
(see the conservation eq. (\ref{eq:2nd}) in Appendix \ref{Sec:Mapping})
which converges for $\tau(q_0,z_0)$ and $  \Delta(q_0,z_0)$ because of the assumption of a finite energy initial flow.

The recent numerical evidence by Luo and Hou \cite{Luo2014}, as well as, the analytic contribution by Elgindi \cite{Elgindi2021annals} place a decisive step in the search of finite-time singularities in Euler equations. Both studies regard axisymmetric flows, moreover, Elgindi imposes the extra condition of a null swirl velocity. 
The question on singularities for an arbitrary flow, relaxing the axisymmetric configuration, remains open. Could the singularity survive to small non-axially symmetric perturbations? On the other hand, some recent numerical study by Kerr \cite{kerr2013} for an initial antiparallel vortex configuration discarded the existence of a finite-time singularity.  Could the manifestation of a singular flow depend on the geometry and symmetries of the flow~? It seems plausible, this question deserves a more careful study.

Another question which must be regarded in more detail concerns the differences and similarities with Elgindi's work \cite{Elgindi2021annals}. A major difference is that Elgindi imposes a null swirl velocity, the price to pay is that the set of functions used is not class $C^\infty$. More importantly, the non-local dependence of the velocity field is approximated (the Biot-Savart integral) by a simpler non-locality.  
Elgindi considers non-smooth dependence in the angular variables (Ref. \cite{Elgindi2021annals} uses spherical coordinates instead of cylindrical coordinates used here), such that the radial dependence is a slowly varying variable compared with angular dependence.  Although the treatment of the non-local terms differs in both approaches, the non-smooth dependence in the angular variables appears to be consistent to the assumption of anisotropy discussed in the current paper. Perhaps the anisotropic assumption may be relaxed by treating the non-local effects as Elgindi, as has been sketched in Sec. \ref{Sec:VerticalSpeed}.
Contrarily, for the axisymmetric flow without swirl, Elgindi neglects the advective term, ${\bm v}\cdot {\bm \nabla}$, obtaining a closed system which exhibits a finite-time singularity for the axial vorticity $\omega_\phi$. While, in the current paper the nonlinear hyperbolic character of the equations emerges at the origin of the appearance of non-smooth velocity field. Finally, the secondary singularity discussed in section \ref{Sec:ExistenceOfFiniteTimeSingularity} seems to be of the same nature as the one found by Elgindi \cite{Elgindi2021annals}.

All these promising results open a possible new endeavor for numerical simulations,  or theoretical analysis based on anisotropic solutions in other simpler geometries, or  the solution of Euler-Leray equations for a more general anisotropic  flow, as well as, including viscosity to the original axisymmetric flow. We shall follow this line of research in a future publication.

\acknowledgements{ Part of this work was done while the author was at the Universidad Adolfo Ibanez 2009-2021. The author acknowledges  P. Clavin,  F. Mora and the anonymous referees for their valuable comments, which lead to significant improvements in the presentation of the current version of the paper.
The author express his gratitude to R. Baquero, M. Le Berre, and Y. Pomeau for their constant interest in this work as well as for numerous discussions.  
This work was supported by FONDECYT under Grant N$^\circ$  1181382.
} 

\section{Appendix: The solution of (\ref{eq:Hamiltonq}-\ref{eq:w}) by means of a dynamical systems approach.}\label{Sec:Mapping}


The four dimensional dynamical system (\ref{eq:Hamiltonq}-\ref{eq:w}) is formally reduced to a two dimensional dynamical system, because of the existence of two constants of motion: 
   \begin{eqnarray}
 \frac{d}{dt} \left( q w\right)  & = &0    , \nonumber \label{eq:qw}\\
\frac{d}{dt} \left(q( J^2 + w^2)\right)    & = & 0 . \nonumber \label{eq:2nd}
    \end{eqnarray} 
    Thus, by virtue of  these conservation laws,
    one  can compute explicitly
       \begin{eqnarray}
 w(q)& = &   \frac{q_0 }{q }  w_0(q_0,z_0) , \nonumber \\
J^2 (q)   & = &  \frac{q _0}{q}  (J(q_0,z_0) ^2 + w_0(q_0,z_0) ^2 )  -   \frac{q^2_0  }{q^2}  w_0(q_0,z_0)^2. \nonumber\\ \label{eq:qJ}
\end{eqnarray} 
    where $J(q_0,z_0)$ and $ w_0(q_0,z_0) $ are initial values at $(q_0,z_0) $.
By solving (\ref{eq:Hamiltonq}) one recovers the time dependent evolution of $q(t)$ already calculated in eq. (\ref{eq:q(t)}), and the sub-sequent time dependent evolution of $J$ (\ref{eq:J(t)}), and $w$ (\ref{eq:w(t)}).   

However we are interested in a more qualitative approach of  the dynamical system (\ref{eq:Hamiltonq}-\ref{eq:w}),   including $z(t)$. The  coordinate variables, $q(t),z(t)$ rule a Hamiltonian dynamics  (\ref{eq:Hamiltonq},\ref{eq:Hamiltonz}) :

\begin{eqnarray}
    \frac{d q  }{d t} & =&  -2  \frac{\partial \psi }{\partial z } 
    , \label{eq:HamiltonqBis} \\
  \frac{d z  }{d t} & =&  2 \frac{\partial \psi }{\partial q } , \label{eq:HamiltonzBis}
      \end{eqnarray}
 for a  Hamiltonian $H= 2 \psi$. Therefore, for a given stream function, $\psi$, the dynamics of $q(t)$ and $z(t)$ follows directly from  Hamilton equations (\ref{eq:HamiltonqBis}) and (\ref{eq:HamiltonzBis}).
      
Nevertheless, the evolution of (\ref{eq:HamiltonqBis},\ref{eq:HamiltonzBis}) is not obvious since $\psi$ depends formally on $J$ by equations (\ref{eq:NewVariables}) and (\ref{eq:qJ}).
From a qualitative point of view one characterizes the mapping of the dynamical system in the $q-z$ plane.

Furthermore, the evolution of  the points in phase space $ (q, z) $ is characterized by:
\begin{enumerate}
\item 
The origin, $q_0=0$ and $z_0=0$, is a fixed point of the dynamics. Moreover, for any $z_0$ and if initially $q_0=0$, then $q(t)=0$ for $t\geq0$. Finally, because the $q$ and $z$ variables rule a Hamiltonian evolution, this is an hyperbolic point.

\item The surrounding region around the rotation axis $q_0=0$ flows according (\ref{eq:HamiltonqBis},\ref{eq:HamiltonzBis}) resulting in a  deformation of the phase space around the rotation axis.

\item Consider the evolution of the set of points $(q_0,z_0)$ which initially belong to the manifold $\Delta(q_0,z_0)=0$, and $\tau(q_0,z_0)>0$. This set of point evolves in time according to   (\ref{eq:qDelta0}).
 The point $(q_0^*,z_0^*) $ reaches the origin $q\to0 $ as $ t \to  t_*=\tau(q_0^*,z_0^*)$, however neighbouring points are excluded from  the axis, for $q>0$. This behavior is at the core of the multi-valued behavior in Eulerian variables $J$ and $w$.

\item The surrounding region of the manifold $ \Delta (q_0, z_0)> 0 $, yields an axial velocity such that $ v_ \phi> 0 $. The divergence free condition implies that $ v_z> 0 $). 

\item The symmetric region such that $ \Delta (q_0, z_0)< 0 $  manifests the opposite previous flow configuration $ v_ \phi <0 $  (and  $ v_z< 0 $). 

\item Finally,   the time evolution  of other points  preserve the area in the phase space accordingly with Hamiltonian dynamics. 
\end{enumerate}

 The dynamical evolution  of the phase space of  equations (\ref{eq:Hamiltonq},\ref{eq:Hamiltonz})  for $ 0 \leq t <t _ * $ is sketched starting with the simple initial condition (\ref{eq:InitTau}) and (\ref{eq:InitDelta}). 
 
\begin{figure}[h]
 \includegraphics[width=0.9\columnwidth]{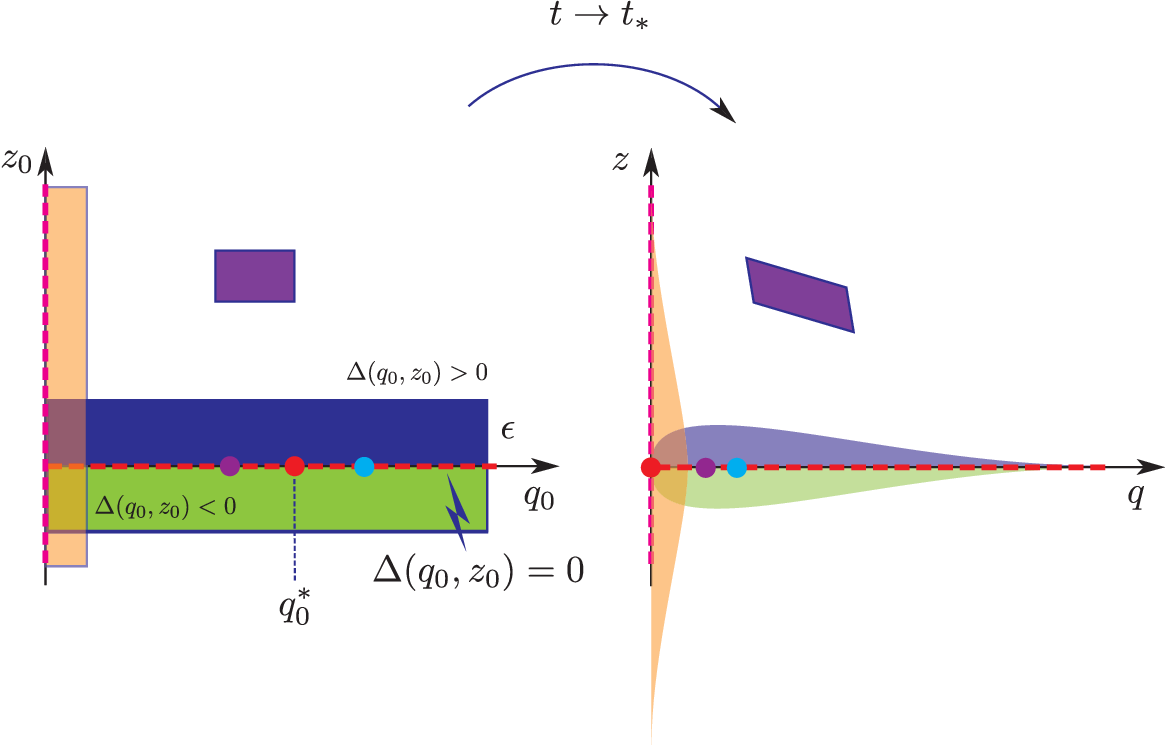}
\caption{  (Left panel) Representative regions in the $ q_0- z_0 $ plane for given initial conditions. The  magenta vertical segmented line  $ q_0 = 0 $ represents the original axis $ r = 0 $. The light orange zone is the neighborhood of the fixed point $ q_0 = 0 $. The red segmented horizontal line $ z_0 = 0 $, represents $ \Delta (q_0, 0) = 0 $ and it corresponds to $ z_0 = 0 $. The red point $ q_0^ * $ corresponds to the minimum of $ \tau (q_0, z_0 = 0) $, and we define: $ t _ * = \tau (q_0 ^*, z_0^* = 0) $. The two extra points (purple, light blue) are located into the neighborhood of $q_0^* $. The blue and green zone belong to the neighborhood of the red segmented line  $ z_0 = 0 $ and $ \Delta (q_0, z_0)> 0 $ in blue and $ \Delta (q_0, z_0) <0 $ into the green region. Finally, the purple zone is an isolated region far from the   $ q_0  $ and $ z_0 $ axes. (Right panel) The resulting mapping representing the evolution of the  initial  space $ (q_0, z_0) $ in to time dependent phase space $(q(t),z(t))$ points through the dynamical system  (\ref{eq:Hamiltonq}-\ref{eq:w}). 
Notice that this picture is only a qualitative sketch. The only certain solutions are the orange points, and the points represented by red, light blue and purple in the axis. } 
\label{fig:mapping}
\end{figure}

 \section{Appendix: Infinite energy blow-up solution}\label{App:InfiniteEnergyExample}
 Setting the particular similarity dependence in the form:
 \begin{eqnarray}
\psi(r,z,t) & = & f(r^2 z,t) , \label{eq:Ansatzf}\\
v_\phi(r,z,t) & = & r w(r^2 z,t) . \label{eq:Ansatzg}
\end{eqnarray}
This {\rm Ansatz} satisfies the boundary conditions at the axis of rotation, indeed  $ v_r = - r \frac{\partial }{\partial \zeta} f(\zeta,t)$ with $\zeta = r^2z$. Therefore, $\lim_{r\to 0} v_r=0$, and, by (\ref{eq:Ansatzg}), $ \lim_{r\to 0} v_\phi=0$. By sustituting  (\ref{eq:Ansatzf}) and  (\ref{eq:Ansatzg}) into  equations  (\ref{eq:AxiSymmetricOmegaBis}) and (\ref{eq:AxiSymmetricGammaBis}) one gets 

\begin{eqnarray}
\frac{\partial}{\partial t}   \left(  \frac{\partial f }{\partial \zeta }  \right) &=&\left(  \frac{\partial f }{\partial \zeta }  \right)^2 - w^2,  \label{eq:ft}\\
 \frac{\partial w  }{\partial t} &=& 2 \left(  \frac{\partial f }{\partial \zeta }  \right) w .\label{eq:gt}
 \end{eqnarray}
Therefore,  the final result is a pure time dependent ordinary differential equation for  $Z(\zeta,t)= \frac{\partial }{\partial \zeta} f(\zeta,t) + i w(\zeta,t) $~: $\frac{d Z }{d t}   =Z^2 ,$
in which the only dependence on the coordinate $\zeta = r^2z$ comes from the initial condition:
$$Z(\zeta,t)= \frac{1  }{\tau(\zeta) - i \Delta (\zeta)  - t},$$
which is characterized by the complex number $\tau(\zeta) - i \Delta (\zeta)$ which is directly related to the initial values for $\frac{\partial }{\partial \zeta} f(\zeta,0)$ and $w(\zeta,0)$.

By following the same argument as in Section \ref{Sec:ExistenceOfFiniteTimeSingularity}, the function shows a finite-time singularity such that if $\zeta_* \in R \quad / \quad \Delta(\zeta_*)=0\quad  \&\quad  \tau(\zeta_*)= t_* >0 $, then  $$ Z(\zeta_*,t) = 1/(t_*-t) .$$

The above example reveals the existence of an underlying  finite-time singularity. The simplified similarity dependence on the variable $\zeta = r^2 z$ brings to light the presence of a special trajectory to be understood. Unfortunately {\rm Ansatz} (\ref{eq:Ansatzf}) and  (\ref{eq:Ansatzg}) is characterized by a motion with infinite energy. Indeed, $ v_z = 2z \frac{\partial f }{\partial \zeta} $, so that
 \begin{widetext}
$$E = \frac{1}{2} \int  {\bm v}^2 \, d^3 x = \pi \int_0^\infty  r dr  \int_{-\infty}^\infty  \left[  \left( \left(\frac{\partial f }{\partial \zeta} \right)^2 +  w^2\right) + \frac{4}{r^6} \zeta^2 \left(\frac{\partial f }{\partial \zeta} \right)^2  \right]  d\zeta \to \infty. $$
 \end{widetext}


\begin{thebibliography}{43}%
\makeatletter
\providecommand \@ifxundefined [1]{%
 \@ifx{#1\undefined}
}%
\providecommand \@ifnum [1]{%
 \ifnum #1\expandafter \@firstoftwo
 \else \expandafter \@secondoftwo
 \fi
}%
\providecommand \@ifx [1]{%
 \ifx #1\expandafter \@firstoftwo
 \else \expandafter \@secondoftwo
 \fi
}%
\providecommand \natexlab [1]{#1}%
\providecommand \enquote  [1]{``#1''}%
\providecommand \bibnamefont  [1]{#1}%
\providecommand \bibfnamefont [1]{#1}%
\providecommand \citenamefont [1]{#1}%
\providecommand \href@noop [0]{\@secondoftwo}%
\providecommand \href [0]{\begingroup \@sanitize@url \@href}%
\providecommand \@href[1]{\@@startlink{#1}\@@href}%
\providecommand \@@href[1]{\endgroup#1\@@endlink}%
\providecommand \@sanitize@url [0]{\catcode `\\12\catcode `\$12\catcode
  `\&12\catcode `\#12\catcode `\^12\catcode `\_12\catcode `\%12\relax}%
\providecommand \@@startlink[1]{}%
\providecommand \@@endlink[0]{}%
\providecommand \url  [0]{\begingroup\@sanitize@url \@url }%
\providecommand \@url [1]{\endgroup\@href {#1}{\urlprefix }}%
\providecommand \urlprefix  [0]{URL }%
\providecommand \Eprint [0]{\href }%
\providecommand \doibase [0]{http://dx.doi.org/}%
\providecommand \selectlanguage [0]{\@gobble}%
\providecommand \bibinfo  [0]{\@secondoftwo}%
\providecommand \bibfield  [0]{\@secondoftwo}%
\providecommand \translation [1]{[#1]}%
\providecommand \BibitemOpen [0]{}%
\providecommand \bibitemStop [0]{}%
\providecommand \bibitemNoStop [0]{.\EOS\space}%
\providecommand \EOS [0]{\spacefactor3000\relax}%
\providecommand \BibitemShut  [1]{\csname bibitem#1\endcsname}%
\let\auto@bib@innerbib\@empty
\bibitem [{\citenamefont {Lichtenstein}(1925)}]{Lichtenstein}%
  \BibitemOpen
  \bibfield  {author} {\bibinfo {author} {\bibfnamefont {Leon}\ \bibnamefont
  {Lichtenstein}},\ }\bibfield  {title} {\enquote {\bibinfo {title} {{\"U}ber
  einige {E}xistenzprobleme der {H}ydrodynamik.}}\ }\href@noop {} {\bibfield
  {journal} {\bibinfo  {journal} {Mat. Zeit. Phys.}\ }\textbf {\bibinfo
  {volume} {23}},\ \bibinfo {pages} {89--154} (\bibinfo {year}
  {1925})}\BibitemShut {NoStop}%
\bibitem [{\citenamefont {Gunther}(1927)}]{Gunther}%
  \BibitemOpen
  \bibfield  {author} {\bibinfo {author} {\bibfnamefont {N.}~\bibnamefont
  {Gunther}},\ }\bibfield  {title} {\enquote {\bibinfo {title} {On the motion
  of fluid in a moving container.}}\ }\href@noop {} {\bibfield  {journal}
  {\bibinfo  {journal} {Izvestia Akad. Nauk USSR, Ser. Fiz. Mat.}\ }\textbf
  {\bibinfo {volume} {20}},\ \bibinfo {pages} {1323--1348} (\bibinfo {year}
  {1927})}\BibitemShut {NoStop}%
\bibitem [{\citenamefont {Leray}(1934)}]{leray1934}%
  \BibitemOpen
  \bibfield  {author} {\bibinfo {author} {\bibfnamefont {Jean}\ \bibnamefont
  {Leray}},\ }\bibfield  {title} {\enquote {\bibinfo {title} {Sur le mouvement
  d'un liquide visqueux emplissant l'espace},}\ }\href {\doibase
  10.1007/BF02547354} {\bibfield  {journal} {\bibinfo  {journal} {Acta Math.}\
  }\textbf {\bibinfo {volume} {63}},\ \bibinfo {pages} {193--248} (\bibinfo
  {year} {1934})}\BibitemShut {NoStop}%
\bibitem [{\citenamefont {Pomeau}(1995)}]{yves1995}%
  \BibitemOpen
  \bibfield  {author} {\bibinfo {author} {\bibfnamefont {Y.}~\bibnamefont
  {Pomeau}},\ }\bibfield  {title} {\enquote {\bibinfo {title} {Singularit\'e
  dans l'\'evolution du fluide parfait},}\ }\href@noop {} {\bibfield  {journal}
  {\bibinfo  {journal} {C. R. Acad. Sci. Paris}\ }\textbf {\bibinfo {volume}
  {321}},\ \bibinfo {pages} {407--411} (\bibinfo {year} {1995})}\BibitemShut
  {NoStop}%
\bibitem [{\citenamefont {Pomeau}(2018)}]{yves2018}%
  \BibitemOpen
  \bibfield  {author} {\bibinfo {author} {\bibfnamefont {Yves}\ \bibnamefont
  {Pomeau}},\ }\bibfield  {title} {\enquote {\bibinfo {title} {On the
  self-similar solution to the {E}uler equations for an incompressible fluid in
  three dimensions},}\ }\href@noop {} {\bibfield  {journal} {\bibinfo
  {journal} {Comptes Rendus M\'ecanique}\ }\textbf {\bibinfo {volume} {346}},\
  \bibinfo {pages} {184 -- 197} (\bibinfo {year} {2018})}\BibitemShut {NoStop}%
\bibitem [{\citenamefont {Pomeau}\ \emph {et~al.}(2019)\citenamefont {Pomeau},
  \citenamefont {{Le Berre}},\ and\ \citenamefont {Lehner}}]{yves2019}%
  \BibitemOpen
  \bibfield  {author} {\bibinfo {author} {\bibfnamefont {Y.}~\bibnamefont
  {Pomeau}}, \bibinfo {author} {\bibfnamefont {M.}~\bibnamefont {{Le Berre}}},
  \ and\ \bibinfo {author} {\bibfnamefont {T.}~\bibnamefont {Lehner}},\
  }\bibfield  {title} {\enquote {\bibinfo {title} {A case of strong non
  linearity: intermittency in highly turbulent flows},}\ }\href@noop {}
  {\bibfield  {journal} {\bibinfo  {journal} {C.R. M\'ecanique}\ }\textbf
  {\bibinfo {volume} {347}},\ \bibinfo {pages} {342--356} (\bibinfo {year}
  {2019})}\BibitemShut {NoStop}%
\bibitem [{\citenamefont {Pomeau}\ and\ \citenamefont {{Le
  Berre}}(2019)}]{martine19}%
  \BibitemOpen
  \bibfield  {author} {\bibinfo {author} {\bibfnamefont {Y.}~\bibnamefont
  {Pomeau}}\ and\ \bibinfo {author} {\bibfnamefont {M.}~\bibnamefont {{Le
  Berre}}},\ }\bibfield  {title} {\enquote {\bibinfo {title} {Blowing-up
  solutions of the axisymmetric {E}uler equations for an incompressible
  fluid},}\ }\href@noop {} {\bibfield  {journal} {\bibinfo  {journal}
  {arXiv:1901.09426}\ } (\bibinfo {year} {2019})}\BibitemShut {NoStop}%
\bibitem [{\citenamefont {Chae}(2007{\natexlab{a}})}]{Chae2007a}%
  \BibitemOpen
  \bibfield  {author} {\bibinfo {author} {\bibfnamefont {Dongho}\ \bibnamefont
  {Chae}},\ }\bibfield  {title} {\enquote {\bibinfo {title} {Nonexistence of
  asymptotically self-similar singularities in the {E}uler and the
  {N}avier--{S}tokes equations},}\ }\href@noop {} {\bibfield  {journal}
  {\bibinfo  {journal} {Mathematische Annalen}\ }\textbf {\bibinfo {volume}
  {338}},\ \bibinfo {pages} {435--449} (\bibinfo {year}
  {2007}{\natexlab{a}})}\BibitemShut {NoStop}%
\bibitem [{\citenamefont {Chae}(2007{\natexlab{b}})}]{Chae2007b}%
  \BibitemOpen
  \bibfield  {author} {\bibinfo {author} {\bibfnamefont {Dongho}\ \bibnamefont
  {Chae}},\ }\bibfield  {title} {\enquote {\bibinfo {title} {Nonexistence of
  {S}elf-{S}imilar {S}ingularities for the 3d incompressible {E}uler
  {E}quations},}\ }\href@noop {} {\bibfield  {journal} {\bibinfo  {journal}
  {Communications in Mathematical Physics}\ }\textbf {\bibinfo {volume}
  {273}},\ \bibinfo {pages} {203--215} (\bibinfo {year}
  {2007}{\natexlab{b}})}\BibitemShut {NoStop}%
\bibitem [{\citenamefont {Chae}(2010)}]{Chae2010}%
  \BibitemOpen
  \bibfield  {author} {\bibinfo {author} {\bibfnamefont {Dongho}\ \bibnamefont
  {Chae}},\ }\bibfield  {title} {\enquote {\bibinfo {title} {On the generalized
  self-similar singularities for the {E}uler and the {N}avier-{S}tokes
  equations},}\ }\href {\doibase https://doi.org/10.1016/j.jfa.2010.02.006}
  {\bibfield  {journal} {\bibinfo  {journal} {Journal of Functional Analysis}\
  }\textbf {\bibinfo {volume} {258}},\ \bibinfo {pages} {2865--2883} (\bibinfo
  {year} {2010})}\BibitemShut {NoStop}%
\bibitem [{\citenamefont {Lamb}(1895)}]{lamb1895hydrodynamics}%
  \BibitemOpen
  \bibfield  {author} {\bibinfo {author} {\bibfnamefont {H.}~\bibnamefont
  {Lamb}},\ }\href {https://books.google.cl/books?id=d\_AoAAAAYAAJ} {\emph
  {\bibinfo {title} {Hydrodynamics}}}\ (\bibinfo  {publisher} {University
  Press},\ \bibinfo {year} {1895})\BibitemShut {NoStop}%
\bibitem [{\citenamefont {Saffman}(1995)}]{Saffman1995vortex}%
  \BibitemOpen
  \bibfield  {author} {\bibinfo {author} {\bibfnamefont {P.G.}\ \bibnamefont
  {Saffman}},\ }\href@noop {} {\emph {\bibinfo {title} {Vortex {D}ynamics}}},\
  Cambridge monographs on mechanics and applied mathematics\ (\bibinfo
  {publisher} {Cambridge University Press},\ \bibinfo {year}
  {1995})\BibitemShut {NoStop}%
\bibitem [{\citenamefont {Penrose}(1965)}]{Penrose1965}%
  \BibitemOpen
  \bibfield  {author} {\bibinfo {author} {\bibfnamefont {Roger}\ \bibnamefont
  {Penrose}},\ }\bibfield  {title} {\enquote {\bibinfo {title} {Gravitational
  {C}ollapse and {S}pace-{T}ime {S}ingularities},}\ }\href {\doibase
  10.1103/PhysRevLett.14.57} {\bibfield  {journal} {\bibinfo  {journal} {Phys.
  Rev. Lett.}\ }\textbf {\bibinfo {volume} {14}},\ \bibinfo {pages} {57--59}
  (\bibinfo {year} {1965})}\BibitemShut {NoStop}%
\bibitem [{\citenamefont {Moore}(1979)}]{Moore1979}%
  \BibitemOpen
  \bibfield  {author} {\bibinfo {author} {\bibfnamefont {Derek~William}\
  \bibnamefont {Moore}},\ }\bibfield  {title} {\enquote {\bibinfo {title} {The
  spontaneous appearance of a singularity in the shape of an evolving vortex
  sheet},}\ }\href {\doibase 10.1098/rspa.1979.0009} {\bibfield  {journal}
  {\bibinfo  {journal} {Proceedings of the Royal Society of London. A.
  Mathematical and Physical Sciences}\ }\textbf {\bibinfo {volume} {365}},\
  \bibinfo {pages} {105--119} (\bibinfo {year} {1979})}\BibitemShut {NoStop}%
\bibitem [{\citenamefont {Siggia}(1985)}]{Siggia85}%
  \BibitemOpen
  \bibfield  {author} {\bibinfo {author} {\bibfnamefont {Eric~D.}\ \bibnamefont
  {Siggia}},\ }\bibfield  {title} {\enquote {\bibinfo {title} {Collapse and
  amplification of a vortex filament},}\ }\href@noop {} {\bibfield  {journal}
  {\bibinfo  {journal} {Phys. Fluids}\ }\textbf {\bibinfo {volume} {28}},\
  \bibinfo {pages} {794--805} (\bibinfo {year} {1985})}\BibitemShut {NoStop}%
\bibitem [{\citenamefont {Pumir}\ and\ \citenamefont
  {Siggia}(1987)}]{PumirSiggia87}%
  \BibitemOpen
  \bibfield  {author} {\bibinfo {author} {\bibfnamefont {Alain}\ \bibnamefont
  {Pumir}}\ and\ \bibinfo {author} {\bibfnamefont {Eric~D.}\ \bibnamefont
  {Siggia}},\ }\bibfield  {title} {\enquote {\bibinfo {title} {Vortex dynamics
  and the existence of solutions to the {N}avier-{S}tokes equations},}\
  }\href@noop {} {\bibfield  {journal} {\bibinfo  {journal} {The Physics of
  Fluids}\ }\textbf {\bibinfo {volume} {30}},\ \bibinfo {pages} {1606--1626}
  (\bibinfo {year} {1987})}\BibitemShut {NoStop}%
\bibitem [{\citenamefont {Morf}\ \emph {et~al.}(1980)\citenamefont {Morf},
  \citenamefont {Orszag},\ and\ \citenamefont {Frisch}}]{Orszag1980}%
  \BibitemOpen
  \bibfield  {author} {\bibinfo {author} {\bibfnamefont {Rudolf~H.}\
  \bibnamefont {Morf}}, \bibinfo {author} {\bibfnamefont {Steven~A.}\
  \bibnamefont {Orszag}}, \ and\ \bibinfo {author} {\bibfnamefont {Uriel}\
  \bibnamefont {Frisch}},\ }\bibfield  {title} {\enquote {\bibinfo {title}
  {Spontaneous singularity in three-dimensional inviscid, incompressible
  flow},}\ }\href@noop {} {\bibfield  {journal} {\bibinfo  {journal} {Phys.
  Rev. Lett.}\ }\textbf {\bibinfo {volume} {44}},\ \bibinfo {pages} {572--575}
  (\bibinfo {year} {1980})}\BibitemShut {NoStop}%
\bibitem [{\citenamefont {Brachet}\ \emph {et~al.}(1983)\citenamefont
  {Brachet}, \citenamefont {Meiron}, \citenamefont {Orszag}, \citenamefont
  {Nickel}, \citenamefont {Morf},\ and\ \citenamefont {Frisch}}]{Brachet1983}%
  \BibitemOpen
  \bibfield  {author} {\bibinfo {author} {\bibfnamefont {Marc~E.}\ \bibnamefont
  {Brachet}}, \bibinfo {author} {\bibfnamefont {Daniel~I.}\ \bibnamefont
  {Meiron}}, \bibinfo {author} {\bibfnamefont {Steven~A.}\ \bibnamefont
  {Orszag}}, \bibinfo {author} {\bibfnamefont {B.~G.}\ \bibnamefont {Nickel}},
  \bibinfo {author} {\bibfnamefont {Rudolf~H.}\ \bibnamefont {Morf}}, \ and\
  \bibinfo {author} {\bibfnamefont {Uriel}\ \bibnamefont {Frisch}},\ }\bibfield
   {title} {\enquote {\bibinfo {title} {Small-scale structure of the
  {T}aylor--{G}reen vortex},}\ }\href@noop {} {\bibfield  {journal} {\bibinfo
  {journal} {Journal of Fluid Mechanics}\ }\textbf {\bibinfo {volume} {130}},\
  \bibinfo {pages} {411--452} (\bibinfo {year} {1983})}\BibitemShut {NoStop}%
\bibitem [{\citenamefont {Brachet}\ \emph {et~al.}(1992)\citenamefont
  {Brachet}, \citenamefont {Meneguzzi}, \citenamefont {Vincent}, \citenamefont
  {Politano},\ and\ \citenamefont {Sulem}}]{Brachet92}%
  \BibitemOpen
  \bibfield  {author} {\bibinfo {author} {\bibfnamefont {M.}~\bibnamefont
  {Brachet}}, \bibinfo {author} {\bibfnamefont {M.}~\bibnamefont {Meneguzzi}},
  \bibinfo {author} {\bibfnamefont {A.}~\bibnamefont {Vincent}}, \bibinfo
  {author} {\bibfnamefont {H.}~\bibnamefont {Politano}}, \ and\ \bibinfo
  {author} {\bibfnamefont {P.L.}\ \bibnamefont {Sulem}},\ }\bibfield  {title}
  {\enquote {\bibinfo {title} {Numerical evidence of smooth self-similar
  dynamics for three dimensional ideal flows},}\ }\href@noop {} {\bibfield
  {journal} {\bibinfo  {journal} {Phys. Fluids A}\ }\textbf {\bibinfo {volume}
  {4}},\ \bibinfo {pages} {2845} (\bibinfo {year} {1992})}\BibitemShut
  {NoStop}%
\bibitem [{\citenamefont {Kerr}(1993)}]{Kerr93}%
  \BibitemOpen
  \bibfield  {author} {\bibinfo {author} {\bibfnamefont {Robert~M.}\
  \bibnamefont {Kerr}},\ }\bibfield  {title} {\enquote {\bibinfo {title}
  {Evidence for a singularity of the three dimensional incompressible {E}uler
  equation},}\ }\href@noop {} {\bibfield  {journal} {\bibinfo  {journal} {Phys.
  Fluids A}\ }\textbf {\bibinfo {volume} {5}},\ \bibinfo {pages} {1725}
  (\bibinfo {year} {1993})}\BibitemShut {NoStop}%
\bibitem [{\citenamefont {Kerr}(2005)}]{Kerr2005}%
  \BibitemOpen
  \bibfield  {author} {\bibinfo {author} {\bibfnamefont {Robert~M.}\
  \bibnamefont {Kerr}},\ }\bibfield  {title} {\enquote {\bibinfo {title}
  {Velocity and scaling of collapsing {E}uler vortices},}\ }\href@noop {}
  {\bibfield  {journal} {\bibinfo  {journal} {Physics of Fluids}\ }\textbf
  {\bibinfo {volume} {17}},\ \bibinfo {pages} {075103} (\bibinfo {year}
  {2005})}\BibitemShut {NoStop}%
\bibitem [{\citenamefont {Boratav}\ and\ \citenamefont
  {Pelz}(1994)}]{BoratavPelz94}%
  \BibitemOpen
  \bibfield  {author} {\bibinfo {author} {\bibfnamefont {Olu\c s~N.}\
  \bibnamefont {Boratav}}\ and\ \bibinfo {author} {\bibfnamefont {Richard~B.}\
  \bibnamefont {Pelz}},\ }\bibfield  {title} {\enquote {\bibinfo {title}
  {Direct numerical simulation of transition to turbulence from a high symmetry
  initial condition},}\ }\href {\doibase 10.1063/1.868166} {\bibfield
  {journal} {\bibinfo  {journal} {Physics of Fluids}\ }\textbf {\bibinfo
  {volume} {6}},\ \bibinfo {pages} {2757--2784} (\bibinfo {year}
  {1994})}\BibitemShut {NoStop}%
\bibitem [{\citenamefont {Gibbon}(2008)}]{Gibbon08}%
  \BibitemOpen
  \bibfield  {author} {\bibinfo {author} {\bibfnamefont {J.D.}\ \bibnamefont
  {Gibbon}},\ }\bibfield  {title} {\enquote {\bibinfo {title} {The
  three-dimensional {E}uler equations: {W}here do we stand?}}\ }\href@noop {}
  {\bibfield  {journal} {\bibinfo  {journal} {Physica D}\ }\textbf {\bibinfo
  {volume} {237}},\ \bibinfo {pages} {1894--1904} (\bibinfo {year}
  {2008})}\BibitemShut {NoStop}%
\bibitem [{\citenamefont {Luo}\ and\ \citenamefont {Hou}(2014)}]{Luo2014}%
  \BibitemOpen
  \bibfield  {author} {\bibinfo {author} {\bibfnamefont {Guo}\ \bibnamefont
  {Luo}}\ and\ \bibinfo {author} {\bibfnamefont {Thomas~Y.}\ \bibnamefont
  {Hou}},\ }\bibfield  {title} {\enquote {\bibinfo {title} {Potentially
  singular solutions of the 3d axisymmetric {E}uler equations},}\ }\href
  {\doibase 10.1073/pnas.1405238111} {\bibfield  {journal} {\bibinfo  {journal}
  {Proceedings of the National Academy of Sciences}\ }\textbf {\bibinfo
  {volume} {111}},\ \bibinfo {pages} {12968--12973} (\bibinfo {year}
  {2014})}\BibitemShut {NoStop}%
\bibitem [{\citenamefont {Barkley}(2020)}]{Barkley2020}%
  \BibitemOpen
  \bibfield  {author} {\bibinfo {author} {\bibfnamefont {Dwight}\ \bibnamefont
  {Barkley}},\ }\bibfield  {title} {\enquote {\bibinfo {title} {A fluid
  mechanic's analysis of the teacup singularity},}\ }\href {\doibase
  10.1098/rspa.2020.0348} {\bibfield  {journal} {\bibinfo  {journal}
  {Proceedings of the Royal Society A: Mathematical, Physical and Engineering
  Sciences}\ }\textbf {\bibinfo {volume} {476}},\ \bibinfo {pages} {20200348}
  (\bibinfo {year} {2020})}\BibitemShut {NoStop}%
\bibitem [{\citenamefont {Elgindi}\ and\ \citenamefont
  {Jeong}(2019)}]{Elgindi2019}%
  \BibitemOpen
  \bibfield  {author} {\bibinfo {author} {\bibfnamefont {T.M.}\ \bibnamefont
  {Elgindi}}\ and\ \bibinfo {author} {\bibfnamefont {I.-J.}\ \bibnamefont
  {Jeong}},\ }\bibfield  {title} {\enquote {\bibinfo {title} {Finite-{T}ime
  {S}ingularity {F}ormation for {S}trong {S}olutions to the {A}xi-symmetric
  3{D} {E}uler equations},}\ }\href@noop {} {\bibfield  {journal} {\bibinfo
  {journal} {Ann. PDE}\ }\textbf {\bibinfo {volume} {5}},\ \bibinfo {pages}
  {16} (\bibinfo {year} {2019})}\BibitemShut {NoStop}%
\bibitem [{\citenamefont {Elgindi}(2021)}]{Elgindi2021annals}%
  \BibitemOpen
  \bibfield  {author} {\bibinfo {author} {\bibfnamefont {Tarek~M.}\
  \bibnamefont {Elgindi}},\ }\bibfield  {title} {\enquote {\bibinfo {title}
  {Finite-time singularity formation for ${C}^{1,\alpha}$ solutions to the
  incompressible {E}uler equations on $\mathbb{R}^3$.}}\ }\href@noop {}
  {\bibfield  {journal} {\bibinfo  {journal} {Annals of Mathematics}\ }\textbf
  {\bibinfo {volume} {194}},\ \bibinfo {pages} {647--727} (\bibinfo {year}
  {2021})}\BibitemShut {NoStop}%
\bibitem [{\citenamefont {Brenner}\ \emph {et~al.}(2016)\citenamefont
  {Brenner}, \citenamefont {Hormoz},\ and\ \citenamefont
  {Pumir}}]{PumirPRF2016}%
  \BibitemOpen
  \bibfield  {author} {\bibinfo {author} {\bibfnamefont {Michael~P.}\
  \bibnamefont {Brenner}}, \bibinfo {author} {\bibfnamefont {Sahand}\
  \bibnamefont {Hormoz}}, \ and\ \bibinfo {author} {\bibfnamefont {Alain}\
  \bibnamefont {Pumir}},\ }\bibfield  {title} {\enquote {\bibinfo {title}
  {Potential singularity mechanism for the {E}uler equations},}\ }\href
  {\doibase 10.1103/PhysRevFluids.1.084503} {\bibfield  {journal} {\bibinfo
  {journal} {Phys. Rev. Fluids}\ }\textbf {\bibinfo {volume} {1}},\ \bibinfo
  {pages} {084503} (\bibinfo {year} {2016})}\BibitemShut {NoStop}%
\bibitem [{\citenamefont {Constantin}\ \emph {et~al.}(1985)\citenamefont
  {Constantin}, \citenamefont {Lax},\ and\ \citenamefont
  {Majda}}]{ConstantinLaxMajda1985}%
  \BibitemOpen
  \bibfield  {author} {\bibinfo {author} {\bibfnamefont {P.}~\bibnamefont
  {Constantin}}, \bibinfo {author} {\bibfnamefont {P.~D.}\ \bibnamefont {Lax}},
  \ and\ \bibinfo {author} {\bibfnamefont {A.}~\bibnamefont {Majda}},\
  }\bibfield  {title} {\enquote {\bibinfo {title} {A simple one-dimensional
  model for the three-dimensional vorticity equation},}\ }\href@noop {}
  {\bibfield  {journal} {\bibinfo  {journal} {Communications on Pure and
  Applied Mathematics}\ }\textbf {\bibinfo {volume} {38}},\ \bibinfo {pages}
  {715--724} (\bibinfo {year} {1985})}\BibitemShut {NoStop}%
\bibitem [{\citenamefont {De~Gregorio}(1990)}]{DeGregorio1990}%
  \BibitemOpen
  \bibfield  {author} {\bibinfo {author} {\bibfnamefont {S.}~\bibnamefont
  {De~Gregorio}},\ }\bibfield  {title} {\enquote {\bibinfo {title} {On a
  one-dimensional model for the three-dimensional vorticity equation},}\
  }\href@noop {} {\bibfield  {journal} {\bibinfo  {journal} {J. Stat. Phys.}\
  }\textbf {\bibinfo {volume} {59}},\ \bibinfo {pages} {1251--1263} (\bibinfo
  {year} {1990})}\BibitemShut {NoStop}%
\bibitem [{\citenamefont {De~Gregorio}(1996)}]{DeGregorio1996}%
  \BibitemOpen
  \bibfield  {author} {\bibinfo {author} {\bibfnamefont {Salvatore}\
  \bibnamefont {De~Gregorio}},\ }\bibfield  {title} {\enquote {\bibinfo {title}
  {A {P}artial {D}ifferential {E}quation {A}rising in a 1{D} {M}odel for the
  3{D} {V}orticity {E}quation},}\ }\href@noop {} {\bibfield  {journal}
  {\bibinfo  {journal} {Mathematical Methods in the Applied Sciences}\ }\textbf
  {\bibinfo {volume} {19}},\ \bibinfo {pages} {1233--1255} (\bibinfo {year}
  {1996})}\BibitemShut {NoStop}%
\bibitem [{\citenamefont {Josserand}\ \emph {et~al.}(2020)\citenamefont
  {Josserand}, \citenamefont {Pomeau},\ and\ \citenamefont
  {Rica}}]{PRFJoss2020}%
  \BibitemOpen
  \bibfield  {author} {\bibinfo {author} {\bibfnamefont {Christophe}\
  \bibnamefont {Josserand}}, \bibinfo {author} {\bibfnamefont {Yves}\
  \bibnamefont {Pomeau}}, \ and\ \bibinfo {author} {\bibfnamefont {Sergio}\
  \bibnamefont {Rica}},\ }\bibfield  {title} {\enquote {\bibinfo {title}
  {Finite-time localized singularities as a mechanism for turbulent
  dissipation},}\ }\href {\doibase 10.1103/PhysRevFluids.5.054607} {\bibfield
  {journal} {\bibinfo  {journal} {Phys. Rev. Fluids}\ }\textbf {\bibinfo
  {volume} {5}},\ \bibinfo {pages} {054607} (\bibinfo {year}
  {2020})}\BibitemShut {NoStop}%
\bibitem [{\citenamefont {Landau}\ and\ \citenamefont
  {Lifshitz}(1959)}]{landau59}%
  \BibitemOpen
  \bibfield  {author} {\bibinfo {author} {\bibfnamefont {L.~D.}\ \bibnamefont
  {Landau}}\ and\ \bibinfo {author} {\bibfnamefont {E.~M.}\ \bibnamefont
  {Lifshitz}},\ }\href@noop {} {\emph {\bibinfo {title} {Fluid {M}echanics}}}\
  (\bibinfo  {publisher} {Pergamon Press},\ \bibinfo {address} {New York},\
  \bibinfo {year} {1959})\BibitemShut {NoStop}%
\bibitem [{\citenamefont {Majda}\ and\ \citenamefont
  {Bertozzi}(2001)}]{majda_bertozzi_2001}%
  \BibitemOpen
  \bibfield  {author} {\bibinfo {author} {\bibfnamefont {Andrew~J.}\
  \bibnamefont {Majda}}\ and\ \bibinfo {author} {\bibfnamefont {Andrea~L.}\
  \bibnamefont {Bertozzi}},\ }\href {\doibase 10.1017/CBO9780511613203} {\emph
  {\bibinfo {title} {Vorticity and {I}ncompressible {F}low}}},\ Cambridge Texts
  in Applied Mathematics\ (\bibinfo  {publisher} {Cambridge University Press},\
  \bibinfo {year} {2001})\BibitemShut {NoStop}%
\bibitem [{\citenamefont {Gibbon}\ \emph {et~al.}(2003)\citenamefont {Gibbon},
  \citenamefont {Moore},\ and\ \citenamefont {Stuart}}]{Gibbon_2003}%
  \BibitemOpen
  \bibfield  {author} {\bibinfo {author} {\bibfnamefont {J~D}\ \bibnamefont
  {Gibbon}}, \bibinfo {author} {\bibfnamefont {D~R}\ \bibnamefont {Moore}}, \
  and\ \bibinfo {author} {\bibfnamefont {J~T}\ \bibnamefont {Stuart}},\
  }\bibfield  {title} {\enquote {\bibinfo {title} {Exact, infinite energy,
  blow-up solutions of the three-dimensional {E}uler equations},}\ }\href
  {\doibase 10.1088/0951-7715/16/5/315} {\bibfield  {journal} {\bibinfo
  {journal} {Nonlinearity}\ }\textbf {\bibinfo {volume} {16}},\ \bibinfo
  {pages} {1823--1831} (\bibinfo {year} {2003})}\BibitemShut {NoStop}%
\bibitem [{\citenamefont {Ne\v{c}as}\ \emph {et~al.}(1996)\citenamefont
  {Ne\v{c}as}, \citenamefont {R\r{u}\v{z}i\v{c}ka},\ and\ \citenamefont
  {\v{S}ver\'ak}}]{Necas1996}%
  \BibitemOpen
  \bibfield  {author} {\bibinfo {author} {\bibfnamefont {J.}~\bibnamefont
  {Ne\v{c}as}}, \bibinfo {author} {\bibfnamefont {M.}~\bibnamefont
  {R\r{u}\v{z}i\v{c}ka}}, \ and\ \bibinfo {author} {\bibfnamefont
  {V.}~\bibnamefont {\v{S}ver\'ak}},\ }\bibfield  {title} {\enquote {\bibinfo
  {title} {On {L}eray's self-similar solutions of the {N}avier-{S}tokes
  equations},}\ }\href {\doibase 10.1007/BF02551584} {\bibfield  {journal}
  {\bibinfo  {journal} {Acta Math.}\ }\textbf {\bibinfo {volume} {176}},\
  \bibinfo {pages} {283--294} (\bibinfo {year} {1996})}\BibitemShut {NoStop}%
\bibitem [{\citenamefont {Kasner}(1921)}]{Kasner1921c}%
  \BibitemOpen
  \bibfield  {author} {\bibinfo {author} {\bibfnamefont {Edward}\ \bibnamefont
  {Kasner}},\ }\bibfield  {title} {\enquote {\bibinfo {title} {Geometrical
  {T}heorems on {E}instein's {C}osmological {E}quations},}\ }\href@noop {}
  {\bibfield  {journal} {\bibinfo  {journal} {American Journal of Mathematics}\
  }\textbf {\bibinfo {volume} {43}},\ \bibinfo {pages} {217--221} (\bibinfo
  {year} {1921})}\BibitemShut {NoStop}%
\bibitem [{\citenamefont {Grauer}\ and\ \citenamefont
  {Sideris}(1991)}]{GrauerSideris91}%
  \BibitemOpen
  \bibfield  {author} {\bibinfo {author} {\bibfnamefont {Rainer}\ \bibnamefont
  {Grauer}}\ and\ \bibinfo {author} {\bibfnamefont {Thomas~C.}\ \bibnamefont
  {Sideris}},\ }\bibfield  {title} {\enquote {\bibinfo {title} {Numerical
  computation of 3d incompressible ideal fluids with swirl},}\ }\href {\doibase
  10.1103/PhysRevLett.67.3511} {\bibfield  {journal} {\bibinfo  {journal}
  {Phys. Rev. Lett.}\ }\textbf {\bibinfo {volume} {67}},\ \bibinfo {pages}
  {3511--3514} (\bibinfo {year} {1991})}\BibitemShut {NoStop}%
\bibitem [{\citenamefont {Pumir}\ and\ \citenamefont
  {Siggia}(1992)}]{PumirSiggia92}%
  \BibitemOpen
  \bibfield  {author} {\bibinfo {author} {\bibfnamefont {Alain}\ \bibnamefont
  {Pumir}}\ and\ \bibinfo {author} {\bibfnamefont {Eric~D.}\ \bibnamefont
  {Siggia}},\ }\bibfield  {title} {\enquote {\bibinfo {title} {Finite-time
  singularities in the axisymmetric three-dimension {E}uler equations},}\
  }\href {\doibase 10.1103/PhysRevLett.68.1511} {\bibfield  {journal} {\bibinfo
   {journal} {Phys. Rev. Lett.}\ }\textbf {\bibinfo {volume} {68}},\ \bibinfo
  {pages} {1511--1514} (\bibinfo {year} {1992})}\BibitemShut {NoStop}%
\bibitem [{\citenamefont {Tao}(2019)}]{Tao2019}%
  \BibitemOpen
  \bibfield  {author} {\bibinfo {author} {\bibfnamefont {Terence}\ \bibnamefont
  {Tao}},\ }\href {https://terrytao.wordpress.com/tag/tarek-elgindi/} {\enquote
  {\bibinfo {title} {Elgindi's approximation of the {B}iot-{S}avart law},}\ }
  (\bibinfo {year} {2019}),\ \bibinfo {note}
  {https://terrytao.wordpress.com/tag/tarek-elgindi/}\BibitemShut {NoStop}%
\bibitem [{\citenamefont {Ukhovskii}\ and\ \citenamefont
  {Iudovich}(1968)}]{Ukhovskii1968}%
  \BibitemOpen
  \bibfield  {author} {\bibinfo {author} {\bibfnamefont {M.R.}\ \bibnamefont
  {Ukhovskii}}\ and\ \bibinfo {author} {\bibfnamefont {V.I.}\ \bibnamefont
  {Iudovich}},\ }\bibfield  {title} {\enquote {\bibinfo {title} {Axially
  symmetric flows of ideal and viscous fluids filling the whole space},}\
  }\href {\doibase https://doi.org/10.1016/0021-8928(68)90147-0} {\bibfield
  {journal} {\bibinfo  {journal} {Journal of Applied Mathematics and
  Mechanics}\ }\textbf {\bibinfo {volume} {32}},\ \bibinfo {pages} {52--62}
  (\bibinfo {year} {1968})}\BibitemShut {NoStop}%
\bibitem [{\citenamefont {Beale}\ \emph {et~al.}(1984)\citenamefont {Beale},
  \citenamefont {Kato},\ and\ \citenamefont {Majda}}]{beale1984}%
  \BibitemOpen
  \bibfield  {author} {\bibinfo {author} {\bibfnamefont {J.~T.}\ \bibnamefont
  {Beale}}, \bibinfo {author} {\bibfnamefont {T.}~\bibnamefont {Kato}}, \ and\
  \bibinfo {author} {\bibfnamefont {A.}~\bibnamefont {Majda}},\ }\bibfield
  {title} {\enquote {\bibinfo {title} {Remarks on the breakdown of smooth
  solutions for the $3$-d {E}uler equations},}\ }\href@noop {} {\bibfield
  {journal} {\bibinfo  {journal} {Comm. Math. Phys.}\ }\textbf {\bibinfo
  {volume} {94}},\ \bibinfo {pages} {61--66} (\bibinfo {year}
  {1984})}\BibitemShut {NoStop}%
\bibitem [{\citenamefont {Kerr}(2013)}]{kerr2013}%
  \BibitemOpen
  \bibfield  {author} {\bibinfo {author} {\bibfnamefont {Robert~M.}\
  \bibnamefont {Kerr}},\ }\bibfield  {title} {\enquote {\bibinfo {title}
  {Bounds for {E}uler from vorticity moments and line divergence},}\ }\href
  {\doibase 10.1017/jfm.2013.325} {\bibfield  {journal} {\bibinfo  {journal}
  {Journal of Fluid Mechanics}\ }\textbf {\bibinfo {volume} {729}},\ \bibinfo
  {pages} {R2} (\bibinfo {year} {2013})}\BibitemShut {NoStop}%
\end{thebibliography}

%

\end{document}